\documentclass[12pt, draftclsnofoot, onecolumn]{IEEEtran}

\usepackage{amsmath,amsfonts}
\usepackage{amsthm,amssymb}
\usepackage{mathrsfs}
\usepackage{algorithmic}
\usepackage{algorithm}
\usepackage{array}
\usepackage[caption=false,font=footnotesize,labelfont=rm,textfont=rm]{subfig}
\usepackage{textcomp}
\usepackage{stfloats}
\usepackage{url}
\usepackage{verbatim}
\usepackage{graphicx}
\usepackage{epstopdf}
\usepackage{amsmath}
\usepackage{subfig}
\usepackage{mathrsfs}

\usepackage{multirow}
\usepackage{enumitem}
\usepackage{amssymb}
\usepackage{amsmath,lipsum}
\usepackage{cuted}
\usepackage[T1]{fontenc}
\usepackage{float}
\usepackage{cite}

\usepackage{setspace}

\onecolumn
\begin{document}

\title{CoMP Transmission in Downlink NOMA-Based Cellular-Connected UAV Networks}
\markboth{}%
{How to Use the IEEEtran \LaTeX \ Templates}
\author{Hongguang~Sun,~\IEEEmembership{Member,~IEEE,}~Linyi~Zhang, 
~Tony~Q.~S.~Quek,~\IEEEmembership{Fellow,~IEEE,}\\
~Xijun~Wang,~\IEEEmembership{Member,~IEEE,}
and~Yan~Zhang,~\IEEEmembership{Member,~IEEE}
}

%\author{Hongguang Sun, \IEEEmembership{Member, IEEE,} Linyi Zhang, Xijun Wang, \IEEEmembership{Member, IEEE,} Yan Zhang, \IEEEmembership{Member, IEEE,} and Tony Q. S. Quek, \IEEEmembership{Fellow, IEEE}}
\maketitle
\begin{abstract}
In this paper, we study the integration between the coordinated multipoint (CoMP) transmission and the non-orthogonal multiple access (NOMA) in the downlink cellular-connected UAV networks with the coexistence of aerial users (AUs) and terrestrial users (TUs). Based on the comparison of the desired signal strength to the dominant interference strength, the AUs are classified into CoMP-AUs and Non-CoMP AUs, where the former receives transmissions from two cooperative BSs, and constructs two exclusive NOMA clusters with two TUs, respectively. A Non-CoMP AU constructs a NOMA cluster with a TU served by the same BS. By leveraging the tools from stochastic geometry, we propose a novel analytical framework to evaluate the performance of the CoMP-NOMA based cellular-connected UAV network in terms of coverage probability, and average ergodic rate. We reveal the superiority of the proposed CoMP-NOMA scheme by comparing with three benchmark schemes, and further quantify the impacts of key system parameters on the network performance. By harvesting the benefits of both CoMP and NOMA, we prove that the proposed framework can provide reliable connection for AUs by using CoMP and enhance the average ergodic rate through NOMA technique as well.
\end{abstract}
\renewcommand{\thefootnote}{}
\footnotetext{Hongguang Sun, and Linyi Zhang are with the College of Information Engineering, Northwest A$\&$F University, Yangling 712100, China (e-mail: hgsun@nwafu.edu.cn; LinyiZhang2020@163.com;).\\
\indent
Tony Q. S. Quek is with the Information Systems Technology and Design Pillar, Singapore University of Technology and Design, Singapore 487372 (e-mail: tonyquek@sutd.edu.sg).\\
\indent
Xijun Wang is with the School of Electronics and Information Technology, Sun Yat-sen University, Guangzhou 510006, China (email: wangxijun@mail.sysu.edu.cn).\\
\indent
Yan Zhang is with the State Key Laboratory of Integrated Service Networks, Xidian University, Xi’an 710071, China (e-mail: yanzhang@xidian.edu.cn).}
\begin{IEEEkeywords}
Cellular-connected UAV networks, CoMP, NOMA, Stochastic Geometry.
\end{IEEEkeywords}

\section{Introduction}
\IEEEPARstart{R}{ecently}, unmanned aerial vehicle (UAV) has found its various applications in a wide range of areas, such as agriculture, disaster rescue and other civil industries. To support the realization of UAV communications, it's essential to maintain a reliable connection between the UAV and the operator \cite{ref1,ref2,ref3}. Thanks to the pervasive deployment of cellular networks, cellular-connected UAV has been served as a new paradigm to provide ubiquitous connectivity for the newly joined aerial users (AUs) as well as the existing terrestrial users (TUs) \cite{ref4,ref5,ref6,ref7}. \\
\indent
% 蜂窝连接无人机
Unfortunately, there still exists major challenges for achieving satisfactory service quality for AUs. On one hand, existing cellular architectures are primarily designed for TUs, where base stations (BSs) are equipped with down-tilted antennas to improve the desired signal strength and decrease the inter-cell interference (ICI) of TUs. As a result, the high-altitude AUs can only be served by the BS's sidelobe, leading to the poor coverage probability and achievable rate. On the other hand, the integration of AUs into the existing cellular networks also leads to the performance degradation of coexisting TUs due to the spectrum sharing and the resulted extra interference \cite{ref8}. Therefore, it is of great significance to design advanced technologies to enable the harmonious coexistence between AUs and TUs. %The first theoretical investigation considering the coexistence of AUs and TUs into cellular networks can be traced back to \cite{ref9}. 
The authors in \cite{ref10}\cite{ref11} studied the opposite effects of line-of-sight (LoS) transmissions on the incremental received signal power and the extra aggregated interference, which shows that the adversities dominate the benefits. The spectrum sharing between UAV-to-UAV transmissions and uplink cellular transmissions was investigated in \cite{ref12} where the performance for both underlay mode and overlay mode was analyzed. A comprehensive performance analysis framework for the downlink cellular-connected UAV network was proposed in \cite{ref13}, in which the impacts of tilting the UAV antenna, the traffic load, and the network densification on the coverage probability or achievable throughput were evaluated. The directional antennas were equipped by AUs in \cite{CACC} to restrict the number of interfering downlink BSs.\\ %To enhance the performance of AUs and relieve the degradation to the existing TUs, advanced interference management technologies should be designed in order to enable the harmonious coexistence of AUs and TUs.\\
\indent
 Coordinated multipoint (CoMP) transmission technique has been considered as an effective approach to diminish the negative effect of LoS interference. The authors in \cite{ref14} designed the CoMP transmission scheme for UAV BSs to forward signals from TUs to a central processor. The work in \cite{ref15} exploited downlink coherent CoMP transmissions to support static and three-dimensional (3D) mobile AUs, and verified the effect of CoMP in improving the coverage probability. Although CoMP transmissions can boost the network performance, it results in the waste of channel resources, limits the number of users that can be simultaneously served, and deteriorates the spectral efficiency \cite{ref16}\cite{ref17}. For CoMP transmissions in downlink, all associated BSs for CoMP need to allocate the same channel to a cell-edge user and this channel cannot be allocated to other users simultaneously when orthogonal multiple access (OMA) techniques are employed. The network performance is getting even worse with the increasing number of cell-edge users \cite{ref16}.\\
 \indent
 In order to enhance the spectral efficiency, non-orthogonal multiple access (NOMA) has been considered as a promising multiple access technology for the fifth generation (5G) and beyond 5G (B5G) cellular systems \cite{ref18}\cite{ref19}.
 %Compared with OMA, NOMA is capable of serving multiple users simultaneously on the same channel resource by exploiting the successive interference cancellation (SIC) capability. 
 Specifically, to maximize the sum-rate for downlink transmission with the power domain NOMA (PD-NOMA), BS power allocation enables PD-NOMA users to perform the successive interference cancellation (SIC) according to the ascending order of their channel gains \cite{ref20}\footnote{In this work, we consider PD-NOMA which is simply referred to as NOMA in the following statements. }. 
 %An optimal power allocation strategy for downlink NOMA results in low received signal-to-intra-cell-interference ratio for a lower channel gain user (e.g., cell edge user)  who is also more prone to ICI. 
 System-level and link-level simulations in \cite{ref21} indicated clear benefits of NOMA over OMA in terms of overall system throughput as well as individual users' throughput. Although the applications of NOMA has been considered in the uplink cellular-connected UAV networks \cite{ref22,ref23,ref24,ref25}, only few works considered downlink NOMA in such a network scenario. Amongest, the work \cite{ref26} analyzed the outage probability of AU and TU under the downlink NOMA in a single cell network by leveraging the instantaneous channel gain ranking. A robust NOMA scheme has been proposed in \cite{ref27} where the TU and AU are paired for the data link and control link, respectively. The work in \cite{ref28} considered the scenario of two co-channel cells and proposed a cooperative NOMA scheme. However, the impacts of aggregated interference and key system parameters on the performance of NOMA-enabled cellular-connected UAV network have not been investigated. Motivated by this, the work in \cite{ref29} leveraged the tools from stochastic geometry and proposed an analytical framework to evaluate the network performance under downlink NOMA for coexisting AUs and TUs. 
 %where an intercell interference cancellation coordination (ICIC) scheme was further proposed. 
 However, the utilization of NOMA scheme introduces extra inter-NOMA user interference besides the co-channel interference, which may deteriorate the received Signal-to-Interference-plus-Noise-Ratio (SINR) at AUs which are usually served as far users in NOMA clusters. \\
\indent
%  NOMA + CoMP
 To mitigate the severe effect of ICI, improve the whole system spectral efficiency, and satisfy the rate requirements for AUs while reducing deterioration to the TU's performance, the combination of CoMP with NOMA can be served as one of the promising access techniques \cite{ref30}. 
 %By integrating CoMP and NOMA, both the CoMP transmission can be further enhanced, but also a higher spectral efficiency can be achieved especially for the far NOMA users \cite{ref30}. NOMA with superposition coding is employed in CoMP networks which enables multiple users to share the same orthogonal resource block (RB) by leveraging superposition coding at the transmitter and SIC at the receiver. 
 Recently, the integration of joint transmission CoMP (JT-CoMP) has been widely discussed in NOMA-based multicell downlink transmissions. To be specific, the work in \cite{ref31} proposed a joint CoMP C-NOMA for the enhanced cellular system performance, where only TUs were considered by using the Rayleigh channel model. The work in \cite{ref32} studied  a power allocation problem for maximizing the energy efficiency in downlink CoMP systems with NOMA. The work in \cite{ref33} proposed  propose a user grouping and pairing scheme for a CoMP-NOMA-based system. An analytical framework was designed in \cite{ref34} to evaluate the performance of the proposed CoMP-NOMA scheme in the downlink heterogeneous cloud radio access network (H-CRAN) where the average achievable data rate for each NOMA user was derived by adopting the Rayleigh fading channel model. To the best of our knowledge, the integration of CoMP and NOMA has not been investigated in the cellular-connected UAV networks. Different from the ground-to-ground (G2G) transmissions in the traditional terrestrial cellular networks, the air-to-ground (A2G) channels exhibit a rather different behavior experiencing the LoS transmissions which is dependent on the altitude of AUs. As such, the existing CoMP-NOMA schemes cannot be directly applied to the network scenarios incorporating AUs and TUs. Specifically, the CoMP-NOMA scheme should be carefully designed to exploit the asymmetric channel fading for A2G link and G2G link, and an analytical framework should be developed to thoroughly characterize the gain achieved by the CoMP-NOMA scheme in the cellular-connected UAV networks.\\
\indent
Motivated by the aforementioned, in this paper, we study the amalgamation between JT-CoMP and NOMA technologies in the downlink cellular-connected UAV networks to enhance the spectral efficiency of the whole network, and provide a reliable connection for AUs as well. To the best of our knowledge, the application of a joint interference-aware JT-CoMP with NOMA scheme to the cellular-connected UAV network using tools from stochastic geometry has not been investigated. Our main contributions are listed as follows:

\begin{itemize}
    \item
    We investigate the joint efforts of interference-aware JT-CoMP and NOMA in enhancing the performance for downlink transmissions of a cellular-connected UAV network, where AUs that are vulnerable to ICI are prioritized to trigger the JT-CoMP. Specifically, the AUs are divided into two categories, namely, CoMP-AUs and Non-CoMP AUs. A CoMP AU is allowed to receive transmissions from two cooperative BSs, and constructs two exclusive NOMA clusters with two TUs which are served by the two corresponding BSs, respectively. A Non-CoMP AU is served by one BS and constructs a NOMA cluster with a TU served by the same BS.
	\item
    By leveraging the tools from stochastic geometry, we propose a novel analytical framework to evaluate the performance of the CoMP-NOMA based cellular-connected UAV network in terms of coverage probability for AUs and TUs, the average achievable rate for each user in each NOMA cluster, and the spectral efficiency of the network. Due to the existence of LoS and NLoS A2G transmissions, the expression of the corresponding SIR at the CoMP-AU is different from that of either the existing CoMP-OMA and NOMA-Only schemes for cellular-connected UAVs, nor the traditional CoMP-NOMA scheme for the traditional TUs. By using the Cauchy-Schwarz's inequality and Gamma approximations, we derive the computationally tractable expressions for the above performance metrics.
    \item
    We validate the theoretical analysis by using Monte Carlo simulations, and show the superiority of the proposed CoMP-NOMA scheme by comparing with the NOMA-Only, CoMP-OMA and OMA-Only schemes. We further evaluate the impacts on network performance of key system parameters, such as SIR threshold, BS density, AU's altitude, cooperation threshold, and power allocation coefficient allocated to AU and TU within a NOMA cluster. We then provide practical guidelines for an efficient design of the proposed CoMP-NOMA scheme by optimizing the cooperation threshold, AU's altitude and power control coefficients for AU and TU within a NOMA pair.
\end{itemize}

The rest of the paper is organized as follows: Section II details the system model, and the proposed COMP-NOMA framework. Section III derives the relevant distance distributions and the association probabilities. Section IV presents the performance analysis in terms of coverage probability and average ergodic rate. The simulation and analytical results are then provided in Section V, followed by the conclusions drawn in Section VI.

\section{System Model}
\subsection{Network Model}
\indent
In this paper, we consider a downlink cellular-connected UAV network where BSs are distributed according to a homogeneous PPP $\Phi_{\rm B}$ of intensity $\lambda_{\rm b}$ with a fixed height $h$$_{\rm b}$. The spatial locations of TUs and AUs follow two other homogeneous PPPs of intensities $\lambda_{\rm t}$ and $\lambda_{\rm u}$ with fixed heights $h_{\rm t}$ and $h_{\rm u}$, respectively. We assume a fully-loaded network scenario, i.e., $\lambda_{\rm t} \gg \lambda_{\rm b}$, and $\lambda_{\rm u} \gg \lambda_{\rm b}$, such that each BS has at least one TU and one AU to associate with. We consider the random scheduling scheme, and a BS randomly selects an AU and a TU to serve if more than one AU and/or TU are associated with the BS. The BS antenna is assumed to have vertically directional and horizontally omnidirectional radiation pattern, which can be realized by implementing multiple sector antennas. In practice, the BS antennas are usually down-tilted to provide better coverage for TUs. As such, TUs are assumed to be served by BSs via mainlobe with antenna gain $g_{\rm M}$, and AUs above BSs height are assumed to be served by BSs via sidelobe with antenna gain $g_{\rm m}$ \cite{ref13}. Without loss of generality, we focus on the performance of a typical pair of AU and TU. By Slivynak's theorem, the typical AU is assumed to be located at the origin with a fixed altitude of $h_{\rm u}$. In addition, we define $\Delta h_{\rm u} \triangleq |h_{\rm u} - h_{\rm b}|$ as the height difference between a BS and a random AU, and $ \Delta h_{\rm t} \triangleq |h_{\rm t} - h_{\rm b}| $ as the height difference between a BS and a random TU.  
\subsection{Channel Model}
\indent
The channel model consists of large-scale path loss and small-scale fading. For the aerial communication link between the typical AU and the BS, we consider a practical path loss model incorporating both LoS and NLoS transmissions, where a probabilistic function is employed to compute the LoS probability \cite{PACA}:
\begin{equation}
	\label{fun2}
p^{\rm L}(z) = \dfrac{1}{1+C\exp\left(-B\left[\frac{180}{\pi}\arctan(\Delta h_{\rm u}/z)-C\right]\right)},
\end{equation}
where $z$ is the Euclidean horizontal distance between the typical AU and the considered BS, $C$ and $B$ are environment-dependent constants, $\arctan(\Delta h_{\rm u}/z)$ is the elevation angle between an AU and a BS. The NLoS probability between an AU and a BS given by $p^{\rm N}(z) = 1- p^{\rm L}(z)$. Note that $z=\sqrt{r^2-(\Delta h_{\rm u})^2}$, where $r$ is the 3D distance between the AU and the BS. Thus, the LoS probability can also be expressed as a function of $r$, i.e., $p^{\rm L}(r)=p^{\rm L}(z)|_{z=\sqrt{r^2-(\Delta h_{\rm u})^2}}$. Based on the A2G channel model in (\ref{fun2}), we assume that each AU experiences either LoS or NLoS transmission with a BS independently. From the typical AU's perspective, the set of BSs $\Phi_{\rm B}$ can be decomposed into two inhomogeneous PPPs, i.e., $\Phi^{\rm L}_{\rm B}$ and $\Phi^{\rm N}_{\rm B}$, of intensities $\lambda_{\rm B}p^{\rm L}(z)$ and $\lambda_{\rm B}(1-p^{\rm L}(z))$, respectively. For the sake of clarity, we define a BS as a LoS (NLoS) BS, if the typical AU experiences the LoS (NLoS) transmission to the BS.\\
\indent
The large-scale path loss between the typical AU and its associated BS $b_{0}$ is expressed as
\begin{equation}
	\label{fun3}
\zeta_{v}(b_{0})=A_{v}g_{\rm m}r_{{\rm b}_{v},\rm u}^{-\alpha_{v}}, v\in\{\rm L, N\},
\end{equation}
where $v$ denotes the type of A2G links with L and N being short for LoS and NLoS, respectively. The symbol $ A_{v} $ denotes the path loss constant at the reference distance $ d_{i} =1m $ for the type $ v $ link, $\alpha_{v}$ is defined as the path loss exponent for the type $ v $ link, and $g_{\rm m}$ is the antenna sidelobe gain provided by its associated BS. For simplicity, we define $\eta_{v} \triangleq A_{v}g_{\rm m}$, $v\in\{\rm L, N\}$.\\
\indent
For small-scale fading, we adopt the Nakagami-$m$ model with the probability distribution function (PDF) given by $f(\omega) = \dfrac{2m_{v}^{m_{v}}\omega^{2m_{v}-1}}{\Gamma(m_{v})}\exp(-m_{v}\omega^{2})$, where $ m_{v} $, $v\in$\{L, N\} is the fading parameter assumed to be an integer for analytical tractability with $ m_{\rm L}>m_{\rm N} $. Given $\omega\sim\rm Nakagami$$(m_{v})$, it directly follows that the channel power gain $ |\omega|^{2} \sim$ $\rm Gamma$$(m_{v},1/m_{v}) $, where $\rm Gamma$$(K,\Theta) $ is the Gamma distribution with $K$ and $\Theta$ denoting the shape parameter and scale parameter, respectively.\\
\indent
The large-scale path loss between the typical TU and the tagged BS is
\begin{equation}
\label{fun4}
	\zeta_{\rm t}(b)=A_{\rm t}g_{\rm M}r_{\rm b,t}^{-\alpha_{\rm t}},
\end{equation}
where $ A_{\rm t} $ is the attenuation for the terrestrial link, $g_{\rm M}$ is the antenna mainlobe gain provided by its associated BS, and $ \alpha_{t} $ denotes the terrestrial path loss exponent. For simplicity, we define $\eta_{\rm t} \triangleq A_{\rm t}g_{\rm M}$. The small-scale fading between a TU and a BS is exponentially distributed with the unit mean, which corresponds to the Rayleigh fading.
\begin{figure}[t]
	%是可选项 h表示的是here在这里插入，t表示的是在页面的顶部插入
	\centering
	\includegraphics[scale=1]{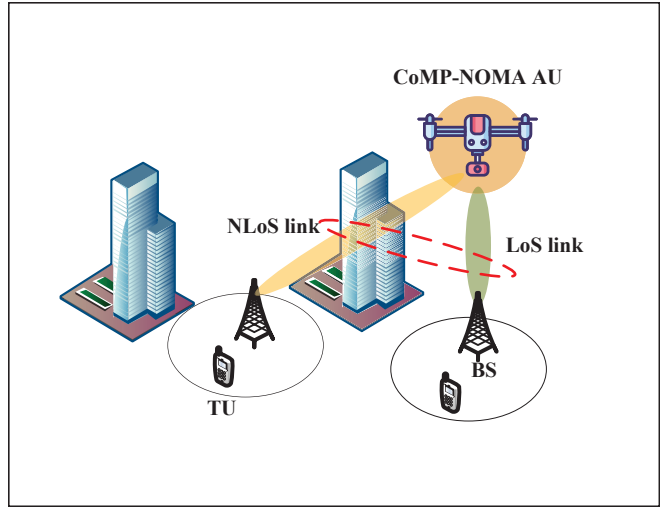}
	\caption{An illustration of the proposed CoMP-NOMA scheme for cellular-connected UAV network.}
	\label{fig1}
\end{figure}
\subsection{User Association and Classification of AUs}
\indent
In this paper, we consider the strongest average received signal strength (RSS) association policy for both AUs and TUs. For a TU, the strongest average RSS association policy is equivalent to the nearest association policy. Therefore, the PDF of the distance between the typical TU and its serving BS is given by $f_R(r)=2\pi\lambda_{b}r\exp(-\pi\lambda_{b}(r^2-\Delta h_{\rm t}^2))$, $r \ge \Delta h_{\rm t}$ \cite{ATAT}. For an AU, due to the LoS/NLoS transmissions, the nearest BS may not be the one providing the strongest average RSS. What's more, we consider the cooperative transmissions for eligible AUs which is served by two BSs. The AUs are classified into Non-CoMP AUs, and CoMP AUs. A CoMP AU is served by the two BSs that provide the first two strongest average RSS, and constructs two exclusive NOMA clusters with two corresponding TUs served by each of the two BSs as shown in Fig. 1. As a result, each CoMP AU is treated as a member in the two corresponding NOMA clusters. For a Non-CoMP AU, it is served by the BS that provides the strongest average RSS, and as a member of a single NOMA cluster. According to the user association policy, the serving BS(s) of a Non-CoMP AU and a CoMP AU are, respectively, selected as follows
\begin{equation}
	\label{fun5}
 b_{0} =\{b_{0} \ | \   \max{\zeta_{v}(b_{v})}, \forall b_{v} \in \Phi_{\rm B}, v\in\{\rm L, \rm N\}\} ,
\end{equation}
\begin{equation}
	\label{fun6}
	\begin{aligned}
		\{b_{0}, b_{1}\} = &\{(b_{0}, b_{1}) \ | \  \max\{(\zeta_{u}(b_{u}), \zeta_{v}(b_{ v}))\}, \forall (b_{u} , b_{v}) \in \Phi_{\rm B}, u,v\in\{\rm L, \rm N\}    \}.
	\end{aligned}
\end{equation}
Similarly, the serving BS of a TU is selected as follows
\begin{equation}
	\label{fun7}
	b_{0} =\{b_{0} \ | \  \max{\zeta_{t}(b)}, \forall b \in \Phi_{\rm B}\}.
\end{equation}
\indent
 We consider an interference-aware AU classification criteria, which is designed based on the ratio of the received signal power from the serving BS, i.e., the BS providing the strongest average RSS, to that from the dominant interfering BS, i.e., the BS providing the second strongest average RSS. Specifically, we define $ \theta $ > 1 as the cooperation threshold. If the aforementioned ratio calculated by an AU is smaller than $ \theta $, the cooperation is activated, and the AU is referred to as the CoMP AU. Otherwise, the AU is only served by the BS providing the strongest average RSS,  referred to as the Non-CoMP AU.\\
\indent
For the typical AU, let $ b_{\rm L_{0}}  $ and $ b_{\rm N_{0}}$ be its nearest LoS BS and NLoS BS, respectively. It is worth noting that the BS providing the strongest average RSS must be either $b_{\rm L_{0}} $ or $ b_{\rm N_{0}} $. Similarly, let $b_{\rm L_{1}} $ and $b_{\rm N_{1}}$ be the second nearest LoS BS and NLoS BS of the typical AU, respectively. Define $\mathcal{B}$ as the serving BS set of the typical AU. Based on the aforementioned criteria of AUs , the serving BS set $\mathcal{B}$ has two possibilities for a Non-CoMP AU, and four possibilities for a CoMP AU. To be specific, if the typical AU is a Non-CoMP AU, $\mathcal{B}=\{b_{\rm L_{0}}\}$ or $ \{b_{\rm N_{0}}\}$ depending on the relation between the RSSs from $b_{\rm L_{0}}$ and $b_{\rm N_{0}}$. If the typical AU is a CoMP AU, $ \mathcal{B}=\{b_{\rm L_{0}},b_{\rm L_{1}}\} $ or $\{b_{\rm N_{0}},b_{\rm N_{1}}\} $ or $\{b_{\rm L_{0}},b_{\rm N_{0}}\} $ or $\{b_{\rm N_{0}},b_{\rm L_{0}}\}$ depending on the rank of the BSs in terms of the RSS. In other words, the serving BSs are the first two BSs providing the strongest average RSSs. The different classifications of AUs are listed in Table I.
\begin{table}
	\begin{center}
		\caption{Classification of AUs}
		\label{tab1}
		\begin{tabular}{|c | c | c | c|}% p指定表列宽度， 内容超过时，会自动产生换行
			\hline	% 表格横线
			Type of AUs & Condition & Type of BSs & Serving BS Set $\mathcal{B}$ \\
			\hline
			\multirow{2}{*}{Non-CoMP}& $ \dfrac{\zeta_{\rm L}}{\zeta_{\rm In}} \ge\theta $ & LoS & $ \{b_{\rm L_{0}}\} $	\\ \cline{2-4}
			& $ \dfrac{\zeta_{\rm N}}{\zeta_{\rm In}} \ge \theta $ &NLoS & $\{ b_{\rm N_{0}}\} $  \\
			\hline
			\multirow{4}{*}{CoMP}& $\dfrac{\zeta_{\rm L_{0}}}{\zeta_{\rm L_{1}}} <\theta, \dfrac{\zeta_{\rm L_{1}}}{\zeta_{\rm N_{0}}} >1 $ & LoS, LoS &  $\{ b_{\rm L_{0}},  b_{\rm L_{1}}\}$	\\\cline{2-4}
			
			& $\dfrac{\zeta_{\rm N_{0}}}{\zeta_{\rm N_{1}}} <\theta, \dfrac{\zeta_{\rm N_{1}}}{\zeta_{\rm L_{0}}} >1 $ & NLoS, NLoS &  $ \{b_{\rm N_{0}},  b_{\rm N_{1}}\}$	\\\cline{2-4}
			
			& $\dfrac{\zeta_{\rm L_{0}}}{\zeta_{\rm N_{0}}} <\theta, \dfrac{\zeta_{\rm N_{0}}}{\zeta_{\rm L_{1}}} >1 $ & LoS, NLoS &  $\{ b_{\rm L_{0}},  b_{\rm N_{0}}\}$	\\\cline{2-4}
   
			& $\dfrac{\zeta_{\rm N_{0}}}{\zeta_{\rm L_{0}}} <\theta, \dfrac{\zeta_{\rm L_{0}}}{\zeta_{\rm N_{1}}} >1 $ & LoS, LoS &  $\{b_{\rm N_{0}},  b_{\rm L_{0}}\} $	\\\cline{2-4}
			\hline
		\end{tabular}
	\end{center}
\end{table}
\subsection{NOMA Model}
We focus on a pair of typical AU and TU associating with the tagged BS. It is worth noting that an appropriate design of pairing strategy between AU and TU for NOMA is also important, which definitely will further enhance the NOMA performance. However, this is beyond the scope of this work and left for the future work. We assume that the AU and the TU are, respectively, the far user and the near user with the corresponding power control coefficients being $\rho_{\rm u}$ and $\rho_{\rm t}$, where $ \rho_{\rm u}$+$\rho_{\rm t} $ =1 and $\rho_{\rm u}  > \rho_{\rm t}$. To maximize the received signal power at the typical AU, we consider the maximum ratio transmission (MRT) scheme, where the tagged BS is assumed to have the channel state information (CSI) between the tagged BS and the typical AU. Take a general $v$-th type BS $b_{i}$ as an example. Define $\omega_{i,\rm u}$ and $\omega_{i,\rm t}$ as the Nakagami-$m$ distributed small-scale fading from $b_{i}$ to its associaed AU and TU, respectively. With MRT, the precoder $w_{i}$ of BS $b_{i}$ is set as $ \dfrac{\omega^{*}_{i,\rm u}}{|\omega_{i,\rm u}|} $, where $ \omega^{*}_{i,\rm u} $ represents the complex conjugate of $ \omega_{i,\rm u} $. To be specific, we use '0' and '1' to represent the typical AU and the typical TU, respectively. In addition, we define $b_{0}$ as the tagged BS and $b_{1}$ as the other cooperative BS if the typical AU is served as a CoMP AU. It is worth noting that $b_{0}$ and $b_{1}$ (if any) can be LoS or NLoS BS. \\
\indent
According to the above definition, the superimposed signal transmitted by the tagged BS $b_{0}$ and the cooperative BS $b_{1}$ (if any) is given by
\begin{equation}
	\label{fun8}
	s_{0} = w_{0}\omega_{0,0}\rho_{\rm u}\sqrt{P_{\rm t}}s_{\rm u}+w_{0}\omega_{0,1}\rho_{\rm t}\sqrt{P_{\rm t}}s_{\rm t},
\end{equation}
\begin{equation}
	\label{fun9}
	s_{1} = w_{1}\omega_{1,0}\rho_{\rm u}\sqrt{P_{\rm t}}s_{\rm u}+w_{1}\omega_{1,\rm t}\rho_{\rm t}\sqrt{P_{\rm t}}s_{\rm t},
\end{equation}
where $w_{0}=\dfrac{\omega^{*}_{0,0}}{|\omega_{0,0}|}$ ($w_{1}=\dfrac{\omega^{*}_{1,0}}{|\omega_{1,0}|}$) is the transmit precoder set by the tagged BS $b_0$ (the cooperative BS $b_1$, if any), $\omega_{0,0}$ ($\omega_{0,1}$) denotes the small-scale fading from the tagged BS $b_0$ to the typical AU (typical TU), and $\omega_{1,0}$ ($\omega_{1,\rm t}$) represents the small-scale fading from the cooperative BS $b_1$ to the typical AU (its associated TU, rather than the typical TU). Meanwhile, $s_{\rm u}$ and $s_{\rm t}$ are the information bearing for AU and TU, respectively, with $\mathbb{E}[|{s_{\rm u}}|^{2}] =  \mathbb{E}[|{s_{\rm t}}|^{2}] = 1$.\\
\indent
We first consider the case when the typical AU is a Non-CoMP AU. In this case, the far user, i.e., the AU, decodes its message directly by treating the signal transmitted to the TU as interference, leading to the following SIR
\begin{equation}
	\label{fun10}
	\Upsilon_{\rm u}^{\rm NC} = \dfrac{\rho_{\rm u}P_{\rm t}\zeta_{v}(b_{0})|\widetilde{\omega}_{0,0}|^2}{\rho_{\rm t}P_{\rm t}\zeta_{v}(b_{0})|\widetilde{\omega}_{0,0}|^2 + \sum_{i\in\Phi_{\rm B}\backslash{b_{0}}}P_{\rm t}\zeta_{v}(b_{i})|\widetilde{\omega}_{i,0}|^2},
\end{equation}
where $|\widetilde{\omega}_{0,0}|^2 \triangleq |w_{0}\omega_{0,0}|^2=|\omega_{0,0}|^2$ ($|\widetilde{\omega}_{i,0}|^2 \triangleq |w_{i}\omega_{i,0}|^2=|\omega_{i,0}|^2$) represents the channel power gain between the tagged BS $b_0$ (interfering BS $b_i$) and the typical AU. Note that the first part in the denominator denotes the self-interference from the tagged BS due to the NOMA scheme between the typical AU and TU.\\
\indent
We then consider the case when the typical AU is a CoMP AU. In this case, the SIR at the AU can be expressed as
\begin{equation}
	\label{fun11}
	\Upsilon_{\rm u}^{\rm C} = \dfrac{|\sum_{k=0}^{1}(\rho_{\rm u}P_{\rm t}\zeta_{v}(b_{k}))^{\frac{1}{2}}\widetilde{\omega}_{k,0}|^{2}}{\sum_{k=0}^{1}\rho_{\rm t}P_{\rm t}\zeta_{v}(b_{k})|\widetilde{\omega}_{k,0}|^{2}+\sum_{i\in\Phi\backslash\mathcal{B}}P_{\rm t}\zeta_{v}(b_{i})|\widetilde{\omega}_{i,0}|^{2}},
\end{equation}
where $\mathcal{B}=\{b_0, b_1\}$ denotes the cooperative BS set, $\widetilde{\omega}_{k,0} = w_{k}\omega_{k,0}=|\omega_{k,0}|$, $|\widetilde{\omega}_{k,0}|^2=|\omega_{k,0}|^2$, $k\in\{0,1\}$, and $|\widetilde{\omega}_{i,0}|^2 \triangleq |w_{i}\omega_{i,0}|^2=|\omega_{i,0}|^2$, $i\in\Phi\backslash\mathcal{B}$. The first part of the denominator denotes the interference from the two cooperative BSs due to the NOMA scheme between the typical AU and TU.\\
\indent
We finally derive the SIR expressions of the typical TU when forming a NOMA pair with a Non-CoMP AU and a CoMP AU, respectively. It is worth noting the small-scale fading from the tagged BS to the typical TU is independent of that from the tagged BS to the typical AU. Thus, the received SIR expression of the typical TU is the same, regardless of the type of the typical AU. Note that in this work, we assume perfect SIC at the typical TU, i.e., the message of the AU can be perfectly removed from the superimposed signal, and thus, the received SIR at the typical TU is given by
\begin{equation}
	\label{fun12}
	\Upsilon_{\rm t} = \dfrac{\rho_{\rm t}P_{\rm t}\zeta_{\rm t}(b_{0})|\widetilde\omega_{0,1}|^{2}}{ \sum_{j\in\Phi\backslash{b_{0}}}P_{\rm t}\zeta_{\rm t}(b_{j})|\widetilde\omega_{j,1}|^{2}},
\end{equation}
where we define $|\widetilde{\omega}_{0,1}|^2 \triangleq |w_{0}\omega_{0,1}|^2=|\omega_{0,1}|^2$. With regards to the Rayleigh distribution, we have $|\widetilde{\omega}_{01}|^{2} \sim$ $\rm Exp(1)$.\\
\indent
For brevity of notation, we define the following symbols and functions, which will be used in the following analysis parts: $l_{\rm L\_N}\triangleq\left(\dfrac{\eta_{\rm L}}{\eta_{\rm N}}\right)^{\frac{1}{\alpha_{\rm L}}}(\Delta h_{\rm u})^{\frac{\alpha_{\rm N}}{\alpha_{\rm L}}}$, $l(r)=\sqrt{r^{2}-(\Delta h_{\rm u})^2}$, $d_{\rm L\_N}(r)=\left(\dfrac{\eta_{\rm N}}{\eta_{\rm L}}\right)^{\frac{1}{\alpha_{\rm N}}}r^{\frac{\alpha_{\rm L}}{\alpha_{\rm N}}}$, and $d_{\rm N\_L}(r)=\left(\dfrac{\eta_{\rm L}}{\eta_{\rm N}}\right)^{\frac{1}{\alpha_{\rm L}}}{r^{\frac{\alpha_{\rm N}}{\alpha_{\rm L}}}}$.

\section{Relevant Distance and User Association Analysis}
\indent
To obtain the coverage probability and ergodic rate, in this section, we first derive the association probabilities when the typical AU is served as a Non-CoMP AU and CoMP AU, respectively. Then, the PDFs of the distance between the tagged BS (BSs) and the typical Non-CoMP AU (CoMP AU) are derived.

\subsection{Relevant Distance Distributions}
In this subsection, we derive the distribution of some relevant distances in Lemma 1 and Lemma 2, which will be used when deriving the association probabilities. \\
\indent
\emph{Lemma 1:} The PDF of the distances between the typical Non-CoMP AU and the closest NLoS BS $b_{\rm {N}_{0}}$ and LoS BS $b_{\rm {L}_{0}}$, denoted by $f_{R_{\rm {N_{0}}}}(r)$ and  $f_{R_{\rm {L_0}}}(r)$, respectively, are given by
\begin{equation}
	\label{fun13}
	\begin{aligned}
    f_{R_{\rm N_{0}}}(r)=2\pi\lambda_{b}rp^{\rm N}(r)
    \exp\bigg(-2\pi\lambda_{b}\int_{0}^{l(r)}zp^{\rm N}(z)dz\bigg),
	\end{aligned}
\end{equation}
\begin{equation}
	\label{fun14}
	\begin{aligned}
    f_{R_{\rm L_{0}}}(r)=2\pi\lambda_{b}rp^{\rm L}(r)
    \exp\bigg(-2\pi\lambda_{b}\int_{0}^{l(r)}zp^{\rm L}(z)dz\bigg),
	\end{aligned}
\end{equation}
where $r\ge\Delta h_{\rm u}$,
%$l(r)=\sqrt{r^{2}-(\Delta h_{\rm u})^2}$, 
$p^{\rm N}(r)=1-p^{\rm L}(r)$ with $p^{\rm L}(r)=p^{\rm L}(z)|_{z=\sqrt{r^2-(\Delta h_{\rm u})^2}}$, $\lambda_{b}p^{\rm N}(r)$ and $\lambda_{b}p^{\rm L}(r)$ represent the intensities of the NLoS BS set $\Phi^{\rm N}_{\rm B}$, and LoS BS set $\Phi^{\rm L}_{\rm B}$, respectively. \\
\indent
\emph{Proof:} The results can be proved by a modification of Lemma 1 in \cite{VHetNet} for the cellular-connected UAV network, which is omitted due to space limitation. $\hfill\square$\\
\indent
\emph{Lemma 2:} The joint PDF of the distances between the typical CoMP AU and the two cooperative BSs of the same type, denoted by $f_{R_{\rm N_0},R_{\rm N_1}}(r_{\rm N_0},r_{\rm N_1})$, and $f_{R_{\rm L_0},R_{\rm L_1}}(r_{\rm L_0},r_{\rm L_1})$, and of the different types, denoted by $f_{R_{\rm N_0},R_{\rm L_0}}(r_{\rm N_0},r_{\rm L_0})$, and $f_{R_{\rm L_0},R_{\rm N_0}}(r_{\rm L_0},r_{\rm N_0})$, are given by
\begin{equation} \label{fun15}
    \begin{aligned}
    f_{R_{\rm N_0},R_{\rm N_1}}(r_{\rm N_0},r_{\rm N_1}) =( 2\pi\lambda_{\rm b})^{2}r_{\rm N_0}r_{\rm N_1}p^{\rm N}(r_{\rm N_0})p^{\rm N}(r_{\rm N_1})\exp \biggl(-2\pi\lambda_{\rm b} \int_{0}^{l(r_{\rm N_1})}zp^{\rm N}(z)dz\biggl),  %\quad r_{\rm N_1} > r_{\rm N_0} \ge \Delta h_{\rm u},
    \end{aligned}	
\end{equation}
%\begin{small}
\begin{equation} \label{fun16}
    \begin{aligned}
    f_{R_{\rm L_0},R_{\rm L_1}}(r_{\rm L_0},r_{\rm L_1}) =( 2\pi\lambda_{\rm b})^{2}r_{\rm L_0}r_{\rm L_1}p^{\rm L}(r_{\rm L_0})p^{\rm L}(r_{\rm L_1})\exp\biggl(-2\pi\lambda_{\rm b} \int_{0}^{l(r_{\rm L_1})}zp^{\rm L}(z)dz\biggl), 
    %\quad r_{\rm L_1} > r_{\rm L_0} \ge \Delta h_{\rm u},
    \end{aligned}	
\end{equation}  
%\end{small}
%\begin{small}
\begin{equation} \label{fun17}
    \begin{aligned}
    f_{R_{\rm N_{0}},R_{\rm L_{0}}}(r_{\rm N_{0}},r_{\rm L_{0}}) =( 2\pi\lambda_{\rm b})^{2}r_{\rm N_{0}}r_{\rm L_{0}}p^{\rm N}(r_{\rm N_{0}})p^{\rm L}(r_{\rm L_{0}})\exp \biggl(-2\pi\lambda_{\rm b} \int_{0}^{l(r_{\rm L_{0}})}zp^{\rm L}(z)dz\biggl),
    %\quad r_{\rm L_{0}} > r_{\rm N_{0}} \geq \Delta h_{\rm u},
    \end{aligned}
\end{equation}    
%\end{small}
%\begin{small}
\begin{equation}  \label{fun18}
    \begin{aligned}
    f_{R_{\rm L_{0}},R_{\rm N_{0}}}(r_{\rm L_{0}},r_{\rm N_{0}}) =
     \begin{cases}
f_{R_{\rm L_{0}}}(r_{\rm L_{0}})f_{R_{\rm N_{0}}}(r_{\rm N_{0}}), &{\text{if}\ r_{\rm L_0} \in [\Delta h_{\rm u},l_{\rm L\_N}), r_{\rm N_0}\ge\Delta h_{\rm u}},\\
(2\pi\lambda_{b})^{2}r_{\rm L_{0}}r_{\rm N_{0}}p^{\rm L}(r_{\rm L_{0}})p^{\rm N}(r_{\rm N_{0}})\\
\qquad \times \exp \left(-2\pi\lambda_{b} \int_{0}^{l(r_{\rm L_{0}})}zp^{\rm L}(z)dz\right), &{\text{if}\  r_{\rm L_{0}} \ge l_{\rm L\_N}, r_{\rm N_0} > d_{\rm L\_N}(r_{\rm L_{0}})}, 
     \end{cases},  
     \end{aligned}
\end{equation}    
%\end{small}
%    \end{flalign}	
where $r_{\rm N_1} > r_{\rm N_0} \ge \Delta h_{\rm u}$ in (\ref{fun15}), $r_{\rm L_1} > r_{\rm L_0} \ge \Delta h_{\rm u}$ in (\ref{fun16}), and $r_{\rm L_{0}} > r_{\rm N_{0}} \geq \Delta h_{\rm u}$ in (\ref{fun17}). $f_{R_{\rm N_{0}}}(r_{\rm N_{0}})$ and $f_{R_{\rm L_{0}}}(r_{\rm L_{0}})$  are given by (\ref{fun13}) and (\ref{fun14}), respectively.
\begin{comment}
$l_{\rm L\_N}\triangleq\left(\dfrac{\eta_{\rm L}}{\eta_{\rm N}}\right)^{\frac{1}{\alpha_{\rm L}}}(\Delta h_{\rm u})^{\frac{\alpha_{\rm N}}{\alpha_{\rm L}}}$, $l(r)=\sqrt{r^{2}-(\Delta h_{\rm u})^2}$, $p^{\rm N}(r)=1-p^{\rm L}(r)$ with $p^{\rm L}(r)=p^{\rm L}(z)|_{z=\sqrt{r^2-(\Delta h_{\rm u})^2}}$, $d_{\rm L\_N}(r_{\rm L_{0}})=\left(\dfrac{\eta_{\rm N}}{\eta_{\rm L}}\right)^{\frac{1}{\alpha_{\rm N}}}{r_{\rm L_{0}}^{\frac{\alpha_{\rm L}}{\alpha_{\rm N}}}}$, $d_{\rm N\_L}(r_{\rm N_{0}})=\left(\dfrac{\eta_{\rm L}}{\eta_{\rm N}}\right)^{\frac{1}{\alpha_{\rm L}}}{r_{\rm N_{0}}^{\frac{\alpha_{\rm N}}{\alpha_{\rm L}}}}$,     
\end{comment}

\indent
\emph{Proof:} See Appendix A. $\hfill\square$\\
\indent
It is worth noting that the difference between $f_{R_{\rm N_{0}}, R_{\rm L{0}}}(r_{\rm N_{0}},r_{\rm L_{0}})$, and $f_{R_{\rm L_{0}}, R_{\rm N_{0}}}(r_{\rm L_{0}},r_{\rm N_{0}})$ lies in the fact that the BS providing the strongest average RSS for the typical AU is of the NLoS type in the former, and of the LoS type in the latter.
\subsection{User Association Analysis}
We define $\mathcal{A}_{\rm L_{0}}$ and $\mathcal{A}_{\rm N_{0}}$ as the probabilities when the typical Non-CoMP AU is associated with the nearest LoS BS and NLoS BS, respectively. We further define $\mathcal{A}_{\rm L_{0},L_{1}} $, $ \mathcal{A}_{\rm N_{0},N_{1}} $, $ \mathcal{A}_{\rm L_{0},N_{0}} $, and $ \mathcal{A}_{\rm N_{0},L_{0}} $ as the probabilities when the typical CoMP AU is associated with $ \{b_{\rm L_{0}}, b_{\rm L_{1}}\} $, $ \{b_{\rm N_{0}}, b_{\rm N_{1}}\} $, $ \{b_{\rm L_{0}}, b_{\rm L_{1}}\} $, and $ \{b_{\rm N_{0}}, b_{\rm L_{0}}\} $, respectively. 
%According to the user association rule, we derive the probabilities of the typical Non-CoMP AU associating with the serving BS, and the typical CoMP AU associating with two serving BSs in the following lemma.
\\
\indent
\emph{Lemma 3:} The probabilities of the typical Non-CoMP AU associated with the serving BS are given by (\ref{fun19}), (\ref{fun20}), and of the typical CoMP AU associated with the two serving BSs for the four different cases are given by (\ref{fun21}), (\ref{fun22}), (\ref{fun23}), and (\ref{fun24}), respectively.
\begin{equation}
	\label{fun19}
\begin{aligned}
	\mathcal{A}_{\rm L_{0}}
&=\int_{\Delta h_{\rm u}}^{l_{\rm L\_N}} \int_{\theta^{\frac{1}{\alpha_{\rm L}}}r_{\rm L_{0}}}^{+\infty}f_{R_{\rm L_{0}},R_{\rm L_{1}}}(r_{\rm L_{0}},r_{\rm L_{1}})dr_{\rm L_{1}}dr_{\rm L_{0}}\\
&+\int_{l_{\rm L\_N}}^{+\infty}\int_{\theta^{\frac{1}{\alpha_{\rm L}}}r_{\rm L_{0}}}^{+\infty}f_{R_{\rm L_{0}},R_{\rm L_{1}}}(r_{\rm L_{0}},r_{\rm L_{1}})\exp\bigg(-2\pi\lambda_{b}\int_{0}^{l\big(\theta^{\frac{1}{\alpha_{\rm N}}}d_{\rm L\_N}(r_{\rm L_{0}})\big)}zp^{\rm N}(z)dz\bigg)dr_{\rm L_{1}}dr_{\rm L_{0}},
\end{aligned}
\end{equation}
\begin{equation}	
	\label{fun20}
	\mathcal{A}_{\rm N_{0}} = \int_{\Delta h_{u}}^{+\infty}\int_{\theta^{\frac{1}{\alpha_{\rm N}}}r_{\rm N_{0}}}^{+\infty}\exp\bigg(-2\pi\lambda_{b}\int_{0}^{l\big(\theta^{\frac{1}{\alpha_{\rm L}}}d_{\rm N\_L}(r_{\rm N_{0}})\big)}zp^{\rm N}(z)dz\bigg) f_{R_{\rm N_{0}},R_{\rm N_{1}}}(r_{\rm N_{0}}, \rm r_{N_{1}} )dr_{\rm N_{1}}dr_{\rm N_{0}},
\end{equation}
\begin{equation}
	\label{fun21}
	\begin{aligned}
		\mathcal{A}_{\rm L_{0},L_{1}}
&= \int_{\Delta h_{\rm u}}^{l_{\rm L\_N}}\int_{r_{\rm L_{0}}}^{\theta^{\frac{1}{\alpha_{\rm L}}}r_{\rm L_{0}}}f_{R_{{\rm L}_{0}},R_{{\rm L}_{1}}}(r_{{\rm L}_{0}},r_{{\rm L}_{1}})dr_{\rm L_{1}}dr_{\rm L_{0}}\\
&+\int_{l_{\rm L\_N}}^{+\infty}\int_{r_{\rm L_{0}}}^{\theta^{\frac{1}{\alpha_{\rm L}}}r_{\rm L_{0}}}f_{R_{{\rm L}_{0}},R_{{\rm L}_{1}}}(r_{{\rm L}_{0}},r_{{\rm L}_{1}})\exp\bigg(-2\pi\lambda_{b}\int_{0}^{l\big(\theta^{\frac{1}{\alpha_{\rm N}}}d_{\rm L\_N}(r_{\rm L_{0}})\big)}zp^{\rm N}(z)dz\bigg)dr_{\rm L_{1}}dr_{\rm L_{0}},
	\end{aligned}
\end{equation}
\begin{equation}
	\label{fun22}
	\mathcal{A}_{\rm N_{0},N_{1}} = \int_{\theta^{\frac{1}{\alpha_{\rm L}}}l_{\rm L\_N}}^{+\infty}
	\int_{\theta^{\frac{1}{\alpha_{\rm N}}}\Delta h_{u}}^{d_{\rm L\_N}(r_{\rm L_{0}})}	
	\int_{\frac{1}{\theta}^{\frac{1}{\alpha_{\rm N}}}r_{\rm N_{1}}}^{r_{\rm N_{1}}} f_{R_{\rm N_{0}}, R_{\rm N_{1}}}(r_{\rm N_{0}},r_{\rm N_{1}})f_{R_{\rm L_{0}}}(r_{\rm L_{0}})	dr_{\rm N_{0}}dr_{\rm N_{1}}dr_{\rm L_{0}},
\end{equation}
\begin{equation}
	\label{fun23}
	\begin{aligned}
		\mathcal{A}_{\rm L_{0},N_{0}} &= \int_{\frac{1}{\theta }^{\frac{1}{\alpha_{\rm L}}}l_{\rm L\_N}}^{l_{\rm L\_{N}}}
		\int_{\Delta h_{u}}^{\theta^{\frac{1}{\alpha_{\rm N}}}d_{\rm L\_N}(r_{\rm L_{0}}) }
		\int_{d_{\rm N\_L}(r_{\rm N_{0}})}^{+\infty}
		f_{R_{\rm L_{0}}}(r_{\rm L_{0}})f_{R_{\rm N_{0}}}(r_{\rm N_{0}})f_{R_{\rm L_{1}}}(r_{\rm L_{1}})dr_{\rm L_{1}}dr_{\rm N_{0}}dr_{\rm L_{0}}\\
		&+\int_{l_{\rm L\_{N}}}^{+\infty}
		\int_{d_{\rm L\_N}(r_{\rm L_{0}})}^{\theta^{\frac{1}{\alpha_{\rm N}}}d_{\rm L\_N}(r_{\rm L_{0}})}
		\int_{d_{\rm N\_L}(r_{\rm N_{0}})}^{+\infty}
		f_{R_{\rm L_{0}},R_{\rm N_{0}}}(r_ {\rm L_{0}},r_{\rm N_{0}})f_{R_{\rm L_{1}}}(r_{\rm L_{1}})dr_{\rm L_{1}}dr_{\rm N_{0}}dr_{\rm L_{0}},
	\end{aligned}
\end{equation}
\begin{equation}
	\label{fun24}
	\begin{aligned}
		& \mathcal{A}_{\rm N_{0},L_{0}} = \int_{\Delta h_{\rm u}}^{+\infty}
		\int_{d_{\rm N\_L}(r_{\rm N_{0}})}^{\theta^{\frac{1}{\alpha_{\rm L}}}d_{\rm N\_L}(r_{\rm N_{0}})}
		\int_{d_{\rm L\_N}(r_{\rm L_{0}})}^{+\infty}
		f_{R_{\rm N_{0}},R_{\rm L_{0}}}(r_{\rm N_{0}},r_{\rm L_{0}})f_{R_{\rm N_{1}}}(r_{\rm N_{1}})dr_{\rm N_{1}}dr_{\rm L_{0}}dr_{\rm N_{0}},
	\end{aligned}
\end{equation}
where 
\begin{comment}
$l_{\rm L\_N}\triangleq\left(\dfrac{\eta_{\rm L}}{\eta_{\rm N}}\right)^{\frac{1}{\alpha_{\rm L}}}(\Delta h_{\rm u})^{\frac{\alpha_{\rm N}}{\alpha_{\rm L}}}$, $l(r)=\sqrt{r^{2}-(\Delta h_{\rm u})^2}$, $d_{\rm L\_N}(r_{\rm L_{0}})=\left(\dfrac{\eta_{\rm N}}{\eta_{\rm L}}\right)^{\frac{1}{\alpha_{\rm N}}}{r_{\rm L_{0}}^{\frac{\alpha_{\rm L}}{\alpha_{\rm N}}}}$, $d_{\rm N\_L}(r_{\rm N_{0}})=\left(\dfrac{\eta_{\rm L}}{\eta_{\rm N}}\right)^{\frac{1}{\alpha_{\rm L}}}{r_{\rm N_{0}}^{\frac{\alpha_{\rm N}}{\alpha_{\rm L}}}}$,    
\end{comment}
$f_{R_{\rm N_{1}}}(r)=f_{R_{\rm N_{0}}}(r)$, and $f_{R_{\rm L_{1}}}(r)=f_{R_{\rm L_{0}}}(r)$ are given by (\ref{fun13}) and (\ref{fun14}), respectively. The joint PDFs $f_{R_{\rm N_0},R_{\rm N_1}}(r_{\rm N_0},r_{\rm N_1})$, $f_{R_{\rm L_0},R_{\rm L_1}}(r_{\rm L_0},r_{\rm L_1})$, $f_{R_{\rm N_{0}},R_{\rm L_{0}}}(r_{\rm N_{0}},r_{\rm L_{0}})$, and $f_{R_{\rm L_{0}},R_{\rm N_{0}}}(r_{\rm L_{0}},r_{\rm N_{0}})$, are given by (\ref{fun15}), (\ref{fun16}), (\ref{fun17}), and (\ref{fun18}), respectively.\\
\indent
\emph{Proof:} See Appendix B. $\hfill\square$
%From Lemma 3, we observe that the association probability is dependant on the cooperation threshold $\theta$ and AU's altitude $h_{\rm u}$. Enlarging $\theta$ or $h_{\rm u}$ leads to an increase in the probability of being a CoMP AU, which is also shown in the simulation and numerical result part. This can be explained by the classification rule of AUs illustrated in Section II. C, which is mainly determined by the comparison of received signal power from the two BSs providing the first two strongest RSSs. 

With the above lemmas, we further derive the PDF of the distances between the typical Non-CoMP AU (CoMP AU) and the serving BS (two serving cooperative BSs) given that the typical Non-CoMP AU (CoMP AU) is associated with a LoS BS or NLoS BS (two LoS BSs, two NLoS BSs, or one LoS BS and one NLoS BS), respectively. \\
\indent
\emph{Lemma 4:} Given that the typical Non-CoMP AU is associated with a LoS BS (NLoS BS), the PDF of the distance between the typical Non-CoMP AU and the serving LoS BS (NLoS BS), denoted by  $ f_{\widetilde{R}_{\rm L_{0}}}(r_{\rm L_{0}})\biggl( f_{\widetilde{R}_{\rm N_{0}}}(r_{\rm N_{0}})\biggl)$, is given by
%\begin{small}
 \begin{equation}
 \begin{aligned}
	f_{\widetilde{R}_{\rm L_{0}}}(r_{\rm L_{0}}) = 
        \label{fun25}
        \begin{cases}
		\frac{1}{\mathcal{A}_{\rm L_{0}}}\int_{\theta^{\frac{1}{\alpha_{\rm L}}}r_{\rm L_{0}}}^{\infty}f_{R_{{\rm L}_{0}},R_{{\rm L}_{1}}}(r_{\rm L_{0}},r_{2})dr_{2}, &{\text{if}}\ r_{\rm L_{0}}\in [\Delta h_{\rm u}, l_{\rm L\_N}),\\
		\frac{1}{\mathcal{A}_{\rm L_{0}}}\exp\left(-2\pi\lambda_{b}
        \int_{0}^{l(\widetilde{d}_{\rm L\_N}(r_{\rm L_{0}}))}zp^{\rm N}(z)dz\right)\\
        \qquad \qquad \qquad \times \int_{\theta^{\frac{1}{\alpha_{\rm L}}}r_{\rm L_{0}}}^{\infty}f_{R_{{\rm L}_{0}},R_{{\rm L}_{1}}}(r_{\rm L_{0}},r_{2})dr_{2},
        &{\text{if}\ r_{\rm L_{0}}\ge l_{\rm L\_N}},
	\end{cases}
 \end{aligned}
\end{equation}   
%\end{small}
%\begin{small}
 \begin{equation}
	\label{fun26}
        \begin{aligned}
	f_{\widetilde{R}_{\rm N_{0}}}(r_{\rm N_{0}}) =
	\frac{1}{\mathcal{A}_{\rm N_{0}}}\exp(-2\pi\lambda_{b} \int_{0}^{l\left(\theta^{\frac{1}{\alpha_{\rm L}}}d_{\rm N\_L}(r_{\rm N_{0}})\right)}zp^{\rm L}(z)dz)\int_{\theta^{\frac{1}{\alpha_{\rm L}}}r_{\rm N_{0}}}^{\infty}f_{R_{{\rm N}_{0}},R_{{\rm N}_{1}}}(r_{\rm N_{0}},r_{2})dr_{2}, 
    %r_{\rm N_{0}} \ge \Delta h_{\rm u},         
 \end{aligned}
\end{equation}   
%\end{small}
where $r_{\rm N_{0}} \ge \Delta h_{\rm u}$ in (\ref{fun26}), 
\begin{comment}
$l_{\rm L\_N}\triangleq\left(\dfrac{\eta_{\rm L}}{\eta_{\rm N}}\right)^{\frac{1}{\alpha_{\rm L}}}(\Delta h_{\rm u})^{\frac{\alpha_{\rm N}}{\alpha_{\rm L}}}$, $l(r)=\sqrt{r^{2}-(\Delta h_{\rm u})^2}$, $d_{\rm L\_N}(r_{\rm L_{0}})=\left(\dfrac{\eta_{\rm N}}{\eta_{\rm L}}\right)^{\frac{1}{\alpha_{\rm N}}}{r_{\rm L_{0}}^{\frac{\alpha_{\rm L}}{\alpha_{\rm N}}}}$, $d_{\rm N\_L}(r_{\rm N_{0}})=\left(\dfrac{\eta_{\rm L}}{\eta_{\rm N}}\right)^{\frac{1}{\alpha_{\rm L}}}{r_{\rm N_{0}}^{\frac{\alpha_{\rm N}}{\alpha_{\rm L}}}}$,     
\end{comment}
$f_{R_{\rm N_{0}}}(r_{\rm N_{0}})$ and $f_{R_{\rm L_{0}}}(r_{\rm L_{0}})$ are given by (\ref{fun13}) and (\ref{fun14}), respectively. \\
\indent
\emph{Proof:} See Appendix C. $\hfill\square$\\
\indent
\emph{Lemma 5:}  Given that the typical CoMP AU is associated with two cooperative BSs, i.e., $\mathcal B=\{b_{\rm L_{0}}, b_{\rm L_{1}}\}, \{b_{\rm N_{0}}, b_{\rm N_{1}}\}, \{b_{\rm L_{0}}, b_{\rm N_{0}}\}$, $\{b_{N_{0}}, b_{L_{0}}\}$, the PDFs of the distance between the typical CoMP AU and the corresponding cooperative BSs, denoted by $f_{\widetilde{R}_{\rm L_{0}},\widetilde{R}_{\rm L_{1}}}(r_{\rm L_{0}},r_{\rm L_{1}})$, $f_{\widetilde{R}_{\rm N_{0}},\widetilde{R}_{\rm N_{1}}}(r_{\rm N_{0}},r_{\rm N_{1}})$, $f_{\widetilde{R}_{\rm L_{0}},\widetilde{R}_{\rm N_{0}}}(r_{\rm L_{0}},r_{\rm N_{0}})$, $f_{\widetilde{R}_{\rm N_{0}},\widetilde{R}_{\rm L_{0}}}(r_{\rm N_{0}},r_{\rm L_{0}})$, are given by
%\begin{small}
 \begin{equation}
	\label{fun27}
\begin{aligned}	
f_{\widetilde{R}_{\rm L_{0}},\widetilde{R}_{\rm L_{1}}}(r_{\rm L_{0}},r_{\rm L_{1}}) = \begin{cases}
		\frac{1}{ \mathcal{A}_{\rm L_{0},L_{1}}}f_{R_{\rm L_{0}},R_{\rm L_{1}}}(r_{\rm L_{0}},r_{\rm L_{1}}), \qquad {\text{if}}\ r_{\rm L_{0}}\in [\Delta h_{\rm u}, l_{\rm L\_N}), r_{\rm L_{1}} > r_{\rm L_{0}},\\
	\frac{1}{\mathcal{A}_{\rm L_{0},L_{1}}}f_{R_{\rm L_{0}},R_{\rm L_{1}}}(r_{\rm L_{0}},r_{\rm L_{1}})\\
    \times \exp\left(-2\pi\lambda_{b} \int_{0}^{d_{\rm L\_N}(r_{\rm L_{1}})}zp^{\rm N}(z)dz\right), \quad \text{if}\ r_{\rm L_{1}} > r_{\rm L_{0}}\ge l_{\rm L\_N},
	\end{cases}
\end{aligned}
\end{equation}   
%\end{small}
%\begin{small}
  \begin{equation}
	\label{fun28}
	f_{\widetilde{R}_{\rm N_{0}},\widetilde{R}_{\rm N_{1}}}(r_{\rm N_{0}},r_{\rm N_{1}}) =
	\frac{1}{\mathcal{A}_{\rm N_{0},N_{1}}}f_{R_{\rm N_{0}},R_{\rm N_{1}}}(r_{\rm N_{0}},r_{\rm N_{1}})\exp\bigg(-2\pi\lambda_{b} \int_{0}^{d_{\rm N\_L}(r_{\rm N_{1}})}zp^{\rm N}(z)dz\bigg), 
    %r_{\rm N_{1}} > r_{\rm N_{0}} \ge \Delta h_{\rm u},
\end{equation}  
%\end{small}
%\begin{small}
 \begin{equation}
	\label{fun29}
	f_{\widetilde{R}_{\rm N_{0}},\widetilde{R}_{\rm L_{0}}}(r_{\rm N_{0}},r_{\rm L_{0}}) =
	\frac{1}{\mathcal{A}_{\rm N_{0},L_{0}}}f_{R_{\rm N_{0}},R_{\rm L_{0}}}(r_{\rm N_{0}},r_{\rm L_{0}})\exp\bigg(-2\pi\lambda_{b} \int_{0}^{d_{\rm L\_N}(r_{\rm L_{0}})}zp^{\rm N}(z)dz\bigg), 
    %r_{\rm L_{0}} > r_{\rm N_{0}} \ge \Delta h_{\rm u},
\end{equation}   
%\end{small}
%\begin{small}
 \begin{equation}
	\label{fun30}
	f_{\widetilde{R}_{\rm L_{0}},\widetilde{R}_{\rm N_{0}}}(r_{\rm L_{0}},r_{\rm N_{0}}) = \begin{cases}
		\frac{1}{\mathcal{A}_{\rm L_{0},N_{0}}}	f_{R_{\rm L_{0}},R_{N_{0}}}(r_{\rm L_{0}},r_{\rm N_{0}}), &{\text{if}\ r_{\rm L_0} \in [\Delta h_{\rm u},l_{\rm L\_N}), r_{\rm N_0}\ge\Delta h_{\rm u}},  \\
		\frac{1}{\mathcal{A}_{\rm L_{0},N_{0}}}f_{R_{\rm L_{0}},R_{N_{0}}}(r_{\rm L_{0}},r_{\rm N_{0}})\\
        \times \exp\bigg(-2\pi\lambda_{b} \int_{0}^{d_{\rm N\_L}(r_{\rm N_{1}})}zp^{\rm N}(z)dz\bigg),  &{\text{if}\  r_{\rm L_{0}} \ge l_{\rm L\_N}, r_{\rm N_0} > d_{\rm L\_N}(r_{\rm L_{0}})},
	\end{cases}
\end{equation}   
%\end{small}
where $r_{\rm N_{1}} > r_{\rm N_{0}} \ge \Delta h_{\rm u}$ in (\ref{fun28}), $r_{\rm L_{0}} > r_{\rm N_{0}} \ge \Delta h_{\rm u}$ in (\ref{fun29}),
\begin{comment}
$l_{\rm L\_N}\triangleq\left(\dfrac{\eta_{\rm L}}{\eta_{\rm N}}\right)^{\frac{1}{\alpha_{\rm L}}}(\Delta h_{\rm u})^{\frac{\alpha_{\rm N}}{\alpha_{\rm L}}}$, $l(r)=\sqrt{r^{2}-(\Delta h_{\rm u})^2}$, $d_{\rm L\_N}(r_{\rm L_{0}})=\left(\dfrac{\eta_{\rm N}}{\eta_{\rm L}}\right)^{\frac{1}{\alpha_{\rm N}}}{r_{\rm L_{0}}^{\frac{\alpha_{\rm L}}{\alpha_{\rm N}}}}$, $d_{\rm N\_L}(r_{\rm N_{0}})=\left(\dfrac{\eta_{\rm L}}{\eta_{\rm N}}\right)^{\frac{1}{\alpha_{\rm L}}}{r_{\rm N_{0}}^{\frac{\alpha_{\rm N}}{\alpha_{\rm L}}}}$,     
\end{comment}
$f_{R_{\rm N_0},R_{\rm N_1}}(r_{\rm N_0},r_{\rm N_1})$, $f_{R_{\rm L_0},R_{\rm L_1}}(r_{\rm L_0},r_{\rm L_1})$, $f_{R_{\rm N_{0}},R_{\rm L_{0}}}(r_{\rm N_{0}},r_{\rm L_{0}})$, and $f_{R_{\rm L_{0}},R_{\rm N_{0}}}(r_{\rm L_{0}},r_{\rm N_{0}})$, are given (\ref{fun15}), (\ref{fun16}), (\ref{fun17}), and (\ref{fun18}), respectively. \\
\indent
\emph{Proof.} See Appendix D. $\hfill\square$
%Note that the PDFs of the distance between the typical AU and the corresponding BS(s) derived in Lemma 4 and Lemma 5 are useful in evaluating the coverage probability and average ergodic rate of the typical AU in the next section.  
\section{Performance Analysis}
\indent
In this section, we focus on analyzing the system performance in terms of coverage probabilities of the typical AU and TU, and average ergodic rate for the proposed framework.
\subsection{Coverage Probability}
%In this subsection, we derive the analytical expressions of the coverage probabilities for the typical AU (Non-CoMP AU and CoMP AU) and the typical TU, respectively. 
Since the typical AU is treated as the far user, the coverage probability is defined as the received SIR given by (\ref{fun9}) and (\ref{fun10}) (for the Non-CoMP AU and CoMP AU) is larger than the predefined SIR threshold T. \\
%According to the PDF of the distance(s) between the AU and its serving BS(s) in (\ref{fun25}) and (\ref{fun26}), we obtain the coverage probability of the Non-CoMP AU in Theorem 1.\\
\indent
\emph{Theorem 1:} Conditioned on associating with the closest NLoS BS $b_{\rm N_0}$ and LoS BS $b_{\rm L_0}$, the DL coverage probabilities of the typical Non-CoMP AU adopting NOMA given the SIR threshold $T$, denoted by $\mathbb P_{\rm N_{0}}(T)$ and $\mathbb P_{\rm L_{0}}(T)$ are, respectively, given by
\begin{equation}
	\label{Theorem 1-N0}
	\begin{aligned}
		\mathbb P_{\rm N_{0}}(T) =&\int_{\Delta h_{\rm u}}^{+\infty}\sum_{k=0}^{m_{\rm N}-1}\frac{\left(-s\right)^{k}}{k !} \frac{\partial^{k}}{\partial s^{k}}\mathcal{L}_{I}\left(s\right)|_{s=\frac{m_{\rm N}{T}}{(\rho_{\rm u}-\rho_{\rm t}{T})P_{\rm t}\zeta_{\rm N}(b_0)}} f_{\widetilde{R}_{\rm N_{0}}}(r_{\rm N_{0}})dr_{\rm N_{0}},\\
	\end{aligned}
\end{equation}
\begin{equation}
	\label{Theorem 1-L0}
	\begin{aligned}
\mathbb P_{\rm L_{0}}(T)
&=\int_{\Delta h_{\rm u}}^{l_{\rm  L\_N}} \sum_{k=0}^{m_{\rm L}-1} \frac{\left(-s\right)^{k}}{k !} \frac{\partial^{k}}{\partial s^{k}} \mathcal{\widehat{L}}_{\rm I}\left(s\right)|_{s=\frac{m_{\rm L}{T}}{(\rho_{\rm u}-\rho_{\rm t} T)P_{\rm t}\zeta_{\rm L}(b_0)}} f_{\widetilde{R}_{\rm L_{0}}}(r_{\rm L_{0}})dr_{\rm L_{0}}\\
&+ \int_{l_{\rm L\_N}}^{+\infty}\sum_{k=0}^{m_{\rm L}-1} \frac{\left(-s\right)^{k}}{k !} \frac{\partial^{k}}{\partial s^{k}} \mathcal{\widetilde{L}}_{\rm I}\left(s\right)|_{s=\frac{m_{\rm L}{T}}{(\rho_{\rm u}-\rho_{\rm t} T)P_{\rm t}\zeta_{\rm L}(b_0)}} f_{\widetilde{R}_{\rm L_{0}}}(r_{\rm L_{0}})dr_{\rm L_{0}},
\end{aligned}
\end{equation}
where 
%$l_{\rm L\_N}\triangleq\left(\dfrac{\eta_{\rm L}}{\eta_{\rm N}}\right)^{\frac{1}{\alpha_{\rm L}}}(\Delta h_{\rm u})^{\frac{\alpha_{\rm N}}{\alpha_{\rm L}}}$, 
$f_{\widetilde{R}_{\rm L_{0}}}(r_{\rm L_{0}})$ and $f_{\widetilde{R}_{\rm N_{0}}}(r_{\rm N_{0}})$ are the PDF of the corresponding distances given by (\ref{fun25}) and (\ref{fun26}), and the expressions of Laplace transform in (\ref{Theorem 1-N0}) and (\ref{Theorem 1-L0}) are, respectively, given by
\begin{equation}
	\begin{aligned}
		\label{laplace_nlos}
        \begin{comment}
		\mathcal{L}_{\rm I}(s)=\exp\bigg(&-2\pi\lambda_{b}\int_{l\left(\theta^{\frac{1}{\alpha_{\rm N}}}r_{\rm N_0}\right)}^{+\infty}\left[1-\left(\dfrac{m_{\rm N}}{m_{\rm N}+s\eta_{\rm N}(\sqrt{z^2+\Delta h_{\rm u}^{2}})^{-\alpha_{\rm N}}}\right)^{m_{\rm N}}\right]zp^{\rm N}(z)dz\\
        &-2\pi\lambda_{b}\int_{l\left(\theta^{\frac{1}{\alpha_{\rm L}}}d_{\rm N\_L}(r_{\rm N_{0}})\right)}^{+\infty}\left[1-\left(\dfrac{m_{\rm L}}{m_{\rm L}+s\eta_{\rm L}(\sqrt{z^2+\Delta h_{\rm u}^{2}})^{-\alpha_{\rm L}}}\right)^{m_{\rm L}}\right]zp^{\rm L}(z)dz\bigg),
        \end{comment}
        \begin{small}
          \mathcal{L}_{\rm I}(s)=\exp\biggl(-2\pi\lambda_{b}\bigg(\int_{l\big(\theta^{\frac{1}{\alpha_{\rm N}}}r_{\rm N_0}\big)}^{+\infty}C(s,z,m_{\rm N})zp^{\rm N}(z)dz+\int_{l\big(\theta^{\frac{1}{\alpha_{\rm L}}}d_{\rm N\_L}(r_{\rm N_{0}})\big)}^{+\infty}C(s,z,m_{\rm L})zp^{\rm L}(z)dz\bigg)\biggl),            
        \end{small}
	\end{aligned}	
\end{equation}
\begin{equation}
	 \label{laplace_los-1}
	\begin{aligned}
    \begin{comment}
		\mathcal{\widehat{L}}_{\rm I}(s)=\exp\bigg(&-2\pi\lambda_{b}\int_{0}^{\infty}\left[1-(\dfrac{m_{\rm N}}{m_{N}+s\eta_{\rm N}(\sqrt{z^2+\Delta h_{\rm u}^{2}})^{-\alpha_{\rm N}}})^{m_{\rm N}}\right]zp_{\rm N}(z)dz\\
		&-2\pi\lambda_{b}\int_{l(r_{\rm L_0})}^{\infty}\left[1-(\dfrac{m_{\rm L}}{m_{\rm L}+s\eta_{\rm L}(\sqrt{z^2+\Delta h_{\rm u}^{2}})^{-\alpha_{\rm L}}})^{m_{\rm L}}\right]zp_{\rm L}(z)dz\bigg) ,\\  
    \end{comment}
    \begin{small}
      \mathcal{\widehat{L}}_{\rm I}(s)=\exp\bigg(-2\pi\lambda_{b}\bigg(\int_{0}^{\infty}C(s,z,m_{\rm N})zp^{\rm N}(z)dz+\int_{l(r_{\rm L_0})}^{\infty}C(s,z,m_{\rm L})zp^{\rm L}(z)dz\bigg)\biggl),        
    \end{small}
	\end{aligned}	
\end{equation}
\begin{equation}
	 \label{laplace_los-2}
	\begin{aligned}
    \begin{comment}
		\mathcal{\widetilde{L}}_{\rm I}(s)=\exp\bigg(&-2\pi\lambda_{b}\int_{l(\theta^{\frac{1}{\alpha_{\rm N}}}d_{\rm L\_N}(r_{\rm L_{0}}))}^{+\infty}\left[1-(\dfrac{m_{\rm N}}{m_{\rm N}+s\eta_{\rm N}(\sqrt{z^2+\Delta h_{\rm u}^{2}})^{-\alpha_{\rm N}}})^{m_{\rm N}}\right]zp_{\rm N}(z)dz\\
		&-2\pi\lambda_{b}\int_{l(r_{\rm L_0})}^{\infty}\left[1-(\dfrac{m_{\rm L}}{m_{\rm L}+s\eta_{\rm L}(\sqrt{z^2+\Delta h_{\rm u}^{2}})^{-\alpha_{\rm L}}})^{m_{\rm L}}\right]zp_{\rm L}(z)dz\bigg),	    
    \end{comment}
    \begin{small}
    \mathcal{\widetilde{L}}_{\rm I}(s)=\exp\bigg(-2\pi\lambda_{b}\bigg(\int_{l(\theta^{\frac{1}{\alpha_{\rm N}}}d_{\rm L\_N}(r_{\rm L_{0}}))}^{+\infty}C(s,z,m_{\rm N})zp^{\rm N}(z)dz+\int_{l(r_{\rm L_0})}^{\infty}C(s,z,m_{\rm L})zp^{\rm L}(z)dz\bigg)\biggl),    
    \end{small}
	\end{aligned}	
\end{equation}
where we define $C(s,z,m_{v}) \triangleq 1-\bigg(\dfrac{m_{v}}{m_{v}+s\eta_{v}\big(\sqrt{z^2+\Delta h_{\rm u}^{2}}\big)^{-\alpha_{v}}}\bigg)^{m_{v}}$, for $v \in \{\rm L, N\}$.
%$l(r)=\sqrt{r^{2}-(\Delta h_{\rm u})^2}$, $d_{\rm N\_L}(r_{\rm N_{0}})=\left(\dfrac{\eta_{\rm L}}{\eta_{\rm N}}\right)^{\frac{1}{\alpha_{\rm L}}}{r_{\rm N_{0}}^{\frac{\alpha_{\rm N}}{\alpha_{\rm L}}}}$.
\\
\indent
\emph{Proof:} The results can be proved by modifying the proof of Theorem 1 in \cite{VHetNet} by incorporating the inter-user interference from NOMA for the cellular-connected UAV network, and omitted here due to space limitation. $\hfill\square$\\
\indent
%The Laplace transform expressions in (\ref{laplace_nlos})-(\ref{laplace_los-2}) show that the aggregated interference is determined by the type of transmission between an interfering BS and the typical Non-CoMP AU, which can be LoS or NLoS.\\    
\emph{Theorem 2:} Conditioned on associating with two cooperative BSs for the four different cases, i.e., $ \mathcal{B}=\{b_{\rm L_{0}},b_{\rm L_{1}}\} $, $\{b_{\rm N_{0}},b_{\rm N_{1}}\} $, $\{b_{\rm L_{0}},b_{\rm N_{0}}\} $ and $\{b_{\rm N_{0}},b_{\rm L_{0}}\}$, the coverage probabilities of the typical CoMP AU adopting NOMA given the SIR threshold $T$, are given by (\ref{Cov-L0L1}), (\ref{Cov-N0N1}), (\ref{Cov_L0N0}), and (\ref{Cov_N0L0}), respectively.
% new LL
\begin{equation}
		\label{Cov-L0L1}
		\begin{aligned}
		\mathbb P_{\rm L_{0},L_{1}}(T)
&\approx \int_{\Delta h_{\rm u}}^{l_{\rm L\_N}}\int_{r_{\rm L_{0}}}^{\theta^{\frac{1}{\alpha_{\rm L}}}r_{\rm L_{0}}} \sum_{k=0}^{2 m_{\rm L}-1} \frac{\left(-s\right)^{k}}{k !} \frac{\partial^{k}}{\partial s^{k}} \mathcal{\widehat{L}}_{I}\left(s\right)|_{s=\frac{T}{\left(2 \rho_{\rm u}-\rho_{\rm t}T\right)P_{\rm t}\Theta}}f_{\widetilde{R}_{\rm L_{0}},\widetilde{R}_{\rm L_{1}}}(r_{\rm L_{0}},r_{\rm L_{1}})dr_{\rm L_{1}}dr_{\rm L_{0}}\\
%		\end{aligned}
%  \nonumber
%\end{equation}
%\begin{equation}
%	\label{Cov-L0L1}
%	\begin{aligned}
&+ \int_{l_{\rm L\_N}}^{+\infty}\int_{r_{\rm L_{0}}}^{\theta^{\frac{1}{\alpha_{\rm L}}}r_{\rm L_{0}}} \sum_{k=0}^{2 m_{\rm L}-1} \frac{\left(-s\right)^{k}}{k !} \frac{\partial^{k}}{\partial s^{k}} \mathcal{\widetilde{L}}_{I}\left(s\right)|_{s=\frac{T}{\left(2 \rho_{\rm u}-\rho_{\rm t}T\right)P_{\rm t}\Theta}}f_{\widetilde{R}_{\rm L_{0}},\widetilde{R}_{\rm L_{1}}}(r_{\rm L_{0}},r_{\rm L_{1}})dr_{\rm L_{1}}dr_{\rm L_{0}},
	\end{aligned}
\end{equation}
	\begin{equation}
		\label{Cov-N0N1}
		\begin{aligned}
			\mathbb P_{\rm N_{0},\rm N_{1}}(T)
\approx \int_{\Delta h_{\rm u}}^{+\infty}\int_{r_{\rm N_{0}}}^{\theta^{\frac{1}{\alpha_{\rm N}}}r_{\rm N_{0}}} \sum_{k=0}^{2 m_{\rm N}-1}\frac{(-s)^{k}}{k !}\frac{\partial^{k}}{\partial s^{k}}\mathcal{L}_{I}(s) |_{s=\frac{{T}}{(2\rho_{\rm u}{P_{\rm t}}-\rho_{\rm t}{P_{\rm t}}{T})\Theta}}f_{\widetilde{R}_{\rm N_{0}},\widetilde{R}_{\rm N_{1}}}(r_{\rm N_{0}},r_{\rm N_{1}})dr_{\rm N_{1}}dr_{\rm N_{0}},
		\end{aligned}
	\end{equation}
	\begin{equation}
		\label{Cov_L0N0}
		\begin{aligned}	
			&\mathbb P_{\rm L_{0},N_{0}}(T)\\ & \approx
\int_{\theta^{-\frac{1}{\alpha_{\rm L}}}l_{\rm L\_N}}^{l_{\rm L\_N}} \int_{\Delta h_{\rm u}}^{\theta^{\frac{1}{\alpha_{\rm N}}}d_{\rm L\_N}(r_{\rm L_{0}})}\sum_{k=0}^{m_{\rm L}+m_{\rm N}-1} \frac{\left(-s\right)^{k}}{k !} \frac{\partial^{k}}{\partial s^{k}} \mathcal{\breve{L}}_{I}\left(s\right)|_{s=\frac{T}{\left(2 \rho_{\rm u}-\rho_{\rm t}T\right)P_{\rm t}\Theta}}f_{\widetilde{R}_{\rm L_{0}},\widetilde{R}_{\rm N_{0}}}(r_{\rm L_{0}},r_{\rm N_{0}})dr_{\rm N_{0}}dr_{\rm L_{0}}\\
			&+\int_{l_{\rm L\_N}}^{+\infty}\int_{d_{\rm L\_N}(r_{\rm L_{0}})}^{\theta^{\frac{1}{\alpha_{\rm N}}}d_{\rm L\_N}(r_{\rm L_{0}})}\sum_{k=0}^{m_{\rm L}+m_{\rm N}-1} \frac{\left(-s\right)^{k}}{k !} \frac{\partial^{k}}{\partial s^{k}} \mathcal{\ddot{L}}_{I}\left(s\right)|_{s=\frac{T}{\left(2 \rho_{\rm u}-\rho_{\rm t}T\right)P_{\rm t}\Theta}}f_{\widetilde{R}_{\rm L_{0}},\widetilde{R}_{\rm N_{0}}}(r_{\rm L_{0}},r_{\rm N_{0}})dr_{\rm N_{0}}dr_{\rm L_{0}},
		\end{aligned}
	\end{equation}	
	\begin{equation}
		\label{Cov_N0L0}
		\begin{aligned}	
			&\mathbb P_{\rm N_{0},L_{0}}(T) \\ & = \int_{\Delta h_{\rm u}}^{+\infty}\int_{d_{\rm N\_L}(r_{\rm N_{0}})}^{\theta^{\frac{1}{\alpha_{\rm L}}}d_{\rm N\_L}(r_{\rm N_{0}})}\sum_{k=0}^{m_{\rm L}+m_{\rm N}-1}\frac{(-s)^{k}}{k !}\frac{\partial^{k}}{\partial s^{k}}\mathcal{\overline{L}}_{\rm I}(s) |_{s=\frac{{T}}{(2\rho_{\rm u}{P_{\rm t}}-\rho_{\rm t}{P_{\rm t}}{T})\Theta}}f_{\widetilde{R}_{\rm N_{0}},\widetilde{R}_{\rm L_{0}}}(r_{\rm N_{0}},r_{\rm L_{0}})dr_{\rm L_{0}}dr_{\rm N_{0}},			
		\end{aligned}
	\end{equation}
where the expressions of Laplace transform are given by
%$l_{\rm L\_N}\triangleq\left(\dfrac{\eta_{\rm L}}{\eta_{\rm N}}\right)^{\frac{1}{\alpha_{\rm L}}}(\Delta h_{\rm u})^{\frac{\alpha_{\rm N}}{\alpha_{\rm L}}}$, $l(r)=\sqrt{r^{2}-(\Delta h_{\rm u})^2}$, $d_{\rm L\_N}(x)=\left(\dfrac{\eta_{\rm N}}{\eta_{\rm L}}\right)^{\frac{1}{\alpha_{\rm N}}}{x}^{\frac{\alpha_{\rm L}}{\alpha_{\rm N}}}$, $d_{\rm N\_L}(x)=\left(\dfrac{\eta_{\rm L}}{\eta_{\rm N}}\right)^{\frac{1}{\alpha_{\rm L}}}{x}^{\frac{\alpha_{\rm N}}{\alpha_{\rm L}}}$, and

\begin{equation}
	\begin{aligned}
		\label{laplace-LL-1}
    \begin{comment}
        \mathcal{\widehat{L}}_{\rm I}(s)
        =\exp\biggl(&-2\pi\lambda_{b}\int_{0}^{+\infty}\left[1-(\dfrac{m_{\rm N}}{m_{\rm N}+s\eta_{\rm N}(\sqrt{z^2+\Delta h_{\rm u}^{2}})^{-\alpha_{\rm N}}})^{m_{\rm N}}\right]zp^{\rm N}(z)dz\\
		&-2\pi\lambda_{b}\int_{l(r_{\rm L_{1}})}^{+\infty}\left[1-\left(\dfrac{m_{\rm L}}{m_{\rm L}+s\eta_{\rm L}(\sqrt{z^2+\Delta h_{\rm u}^{2}})^{-\alpha_{\rm L}}}\right)^{m_{\rm  L}}\right]zp^{\rm L}(z)dz\biggl),\\  
    \end{comment}
    \begin{small}
       \mathcal{\widehat{L}}_{\rm I}(s)=\exp\bigg(-2\pi\lambda_{b}\bigg(\int_{0}^{\infty}C(s,z,m_{\rm N})zp^{\rm N}(z)dz+\int_{l(r_{\rm L_1})}^{\infty}C(s,z,m_{\rm L})zp^{\rm L}(z)dz\bigg)\biggl),       
    \end{small}
	\end{aligned}	
\end{equation}
\begin{equation}
	\begin{aligned}
		\label{laplace-LL-2}
    \begin{comment}
        \mathcal{\widetilde{L}}_{\rm I}(s)
        =\exp\bigg(&-2\pi\lambda_{b}\int_{l(d_{\rm L\_N}(r_{\rm L_{1}}))}^{+\infty}
        \left[1-(\dfrac{m_{\rm N}}{m_{\rm N}+s\eta_{\rm N}(\sqrt{z^2+\Delta h_{\rm u}^{2}})^{-\alpha_{\rm N}}})^{m_{\rm N}}\right]zp^{\rm N}(z)dz\\
		&-2\pi\lambda_{b}\int_{l(r_{\rm L_{1}})}^{+\infty}\left[1-\left(\dfrac{m_{\rm L}}{m_{\rm L}+s\eta_{\rm L}(\sqrt{z^2+\Delta h_{\rm u}^{2}})^{-\alpha_{\rm L}}}\right)^{m_{\rm L}}\right]zp^{\rm L}(z)dz\bigg),\\        
    \end{comment}
    \begin{small}
       \mathcal{\widetilde{L}}_{\rm I}(s)=\exp\bigg(-2\pi\lambda_{b}\bigg(\int_{l(d_{\rm L\_N}(r_{\rm L_{1}}))}^{+\infty}C(s,z,m_{\rm N})zp^{\rm N}(z)dz+\int_{l(r_{\rm L_1})}^{\infty}C(s,z,m_{\rm L})zp^{\rm L}(z)dz\bigg)\biggl),       
    \end{small}
	\end{aligned}	
\end{equation}
\begin{equation}
	\begin{aligned}
		\label{laplace-NN}
    \begin{comment}
        \mathcal{L}_{\rm I}(s)
        =\exp\biggl(&-2\pi\lambda_{b}\int_{l(d_{\rm N\_L}(r_{\rm N_{1}}))}^{+\infty}\left[1-(\dfrac{m_{\rm N}}{m_{\rm N}+s\eta_{\rm N}(\sqrt{z^2+\Delta h_{\rm u}^{2}})^{-\alpha_{\rm N}}})^{m_{\rm N}}\right]zp^{\rm N}(z)dz\\
		&-2\pi\lambda_{b}\int_{l(r_{\rm N_{1}})}^{+\infty}\left[1-\left(\dfrac{m_{\rm L}}{m_{\rm L}+s\eta_{\rm L}(\sqrt{z^2+\Delta h_{\rm u}^{2}})^{-\alpha_{\rm L}}}\right)^{m_{\rm L}}\right]zp^{\rm L}(z)dz\biggl),\\        
    \end{comment}
    \begin{small}
       \mathcal{L}_{\rm I}(s)=\exp\bigg(-2\pi\lambda_{b}\bigg(\int_{l(d_{\rm N\_L}(r_{\rm N_{1}}))}^{+\infty}C(s,z,m_{\rm N})zp^{\rm N}(z)dz+\int_{l(r_{\rm N_1})}^{\infty}C(s,z,m_{\rm L})zp^{\rm L}(z)dz\bigg)\biggl),              
    \end{small}
	\end{aligned}	
\end{equation}
\begin{equation}
	\begin{aligned}
		\label{laplace-LN-1}
    \begin{comment}
        \mathcal{\breve{L}}_{\rm I}(s)
        =\exp\biggl(&-2\pi\lambda_{b}\int_{0}^{+\infty}\left[1-(\dfrac{m_{\rm N}}{m_{\rm N}+s\eta_{\rm N}(\sqrt{z^2+\Delta h_{\rm u}^{2}})^{-\alpha_{\rm N}}})^{m_{\rm N}}\right]zp^{\rm N}(z)dz\\
		&-2\pi\lambda_{b}\int_{l\left(d_{\rm N\_L}(r_{\rm N_{0}})\right)}^{+\infty}\left[1-\left(\dfrac{m_{\rm L}}{m_{\rm L}+s\eta_{\rm L}(\sqrt{z^2+\Delta h_{\rm u}^{2}})^{-\alpha_{\rm L}}}\right)^{m_{\rm L}}\right]zp^{\rm L}(z)dz\biggl),\\        
    \end{comment}
    \begin{small}
       \mathcal{\breve{L}}_{\rm I}(s)=\exp\bigg(-2\pi\lambda_{b}\bigg(\int_{0}^{+\infty}C(s,z,m_{\rm N})zp^{\rm N}(z)dz+\int_{l\left(d_{\rm N\_L}(r_{\rm N_{0}})\right)}^{+\infty}C(s,z,m_{\rm L})zp^{\rm L}(z)dz\bigg)\biggl),             
    \end{small}
	\end{aligned}	
\end{equation}
\begin{equation}
	\begin{aligned}
		\label{laplace-LN-2}
    \begin{comment}
        \mathcal{\ddot{L}}_{\rm I}(s)
        =\exp\bigg(&-2\pi\lambda_{b}\int_{l(r_{\rm N_0})}^{+\infty}
        \left[1-(\dfrac{m_{\rm N}}{m_{\rm N}+s\eta_{\rm N}(\sqrt{z^2+\Delta h_{\rm u}^{2}})^{-\alpha_{\rm N}}})^{m_{\rm N}}\right]zp^{\rm N}(z)dz\\
		&-2\pi\lambda_{b}\int_{l\left(d_{\rm N\_L}(r_{\rm N_{0}})\right)}^{+\infty}\left[1-\left(\dfrac{m_{\rm L}}{m_{\rm L}+s\eta_{\rm L}(\sqrt{z^2+\Delta h_{\rm u}^{2}})^{-\alpha_{\rm L}}}\right)^{m_{\rm L}}\right]zp^{\rm L}(z)dz\bigg),\\        
    \end{comment}
    \begin{small}
       \mathcal{\ddot{L}}_{\rm I}(s)=\exp\bigg(-2\pi\lambda_{b}\bigg(\int_{l(r_{\rm N_0})}^{+\infty}C(s,z,m_{\rm N})zp^{\rm N}(z)dz+\int_{l\left(d_{\rm N\_L}(r_{\rm N_{0}})\right)}^{+\infty}C(s,z,m_{\rm L})zp^{\rm L}(z)dz\bigg)\biggl),            
    \end{small}
	\end{aligned}	
\end{equation}
\begin{equation}
	\begin{aligned}
		\label{laplace-NL}
    \begin{comment}
        \mathcal{\overline{L}}_{\rm I}(s)
        =\exp\biggl(&-2\pi\lambda_{b}\int_{l(d_{\rm L\_N}(r_{\rm L_{0}}))}^{+\infty}\left[1-(\dfrac{m_{\rm N}}{m_{\rm N}+s\eta_{\rm N}(\sqrt{z^2+\Delta h_{\rm u}^{2}})^{-\alpha_{\rm N}}})^{m_{\rm N}}\right]zp^{\rm N}(z)dz\\
		&-2\pi\lambda_{b}\int_{l(r_{\rm L_{0}})}^{+\infty}\left[1-\left(\dfrac{m_{\rm L}}{m_{\rm L}+s\eta_{\rm L}(\sqrt{z^2+\Delta h_{\rm u}^{2}})^{-\alpha_{\rm L}}}\right)^{m_{\rm L}}\right]zp^{\rm L}(z)dz\biggl),\\        
    \end{comment}
    \begin{small}
       \mathcal{\overline{L}}_{\rm I}(s)=\exp\bigg(-2\pi\lambda_{b}\bigg(\int_{l(d_{\rm L\_N}(r_{\rm L_{0}}))}^{+\infty}C(s,z,m_{\rm N})zp^{\rm N}(z)dz+\int_{l(r_{\rm L_{0}})}^{+\infty}C(s,z,m_{\rm L})zp^{\rm L}(z)dz\bigg)\biggl),          
    \end{small}
	\end{aligned}	
\end{equation}
where we define $C(s,z,m_{v}) \triangleq 1-\bigg(\dfrac{m_{v}}{m_{v}+s\eta_{v}\big(\sqrt{z^2+\Delta h_{\rm u}^{2}}\big)^{-\alpha_{v}}}\bigg)^{m_{v}}$, for $v \in \{\rm L, N\}$.\\
\indent
\emph{Proof:} See Appendix E. $\hfill\square$\\
\indent
According to the law of total probability, the overall coverage probability of the typical CoMP AU adopting NOMA given the SIR threshold $T$ can be expressed as
\begin{equation}
	\label{fun41}
	\begin{aligned}
		\mathbb P_{\rm u}(T) & = \mathcal{A}_{\rm L_{0}} \mathbb P_{\rm L_{0}}(T)+ A_{\rm N_{0}}	\mathbb P_{\rm N_{0}}(T)+\mathcal{A}_{\rm L_{0},L_{1}}\mathbb P_{\rm L_{0},L_{1}}(T)\\ &+\mathcal{A}_{\rm N_{0},N_{1}}\mathbb P_{\rm N_{0},N_{1}}(T)+\mathcal{A}_{\rm L_{0},N_{0}}\mathbb P_{\rm L_{0},N_{0}}(T)+\mathcal{A}_{\rm N_{0},L_{0}}\mathbb P_{\rm N_{0},L_{0}}(T).
	\end{aligned}
\end{equation}
By substituting the expressions of association probabilities in Lemma 3, and the coverage probabilities given in Theorem 1 and Theorem 2 into (\ref{fun41}), we can derive the coverage probability of the typical AU.\\
\indent
Note that the typical TU is treated as the near user in this work, its coverage probability is defined as the received SIR given by (\ref{fun11}) is larger than the predefined SIR threshold $T$, which is given by the following corollary.\\
\indent
\emph{Corollary 1:} The coverage probability of the typical TU is given by
\begin{equation}
	\begin{aligned}
		\mathbb P_{\rm t}(T) = 2\pi\lambda_{b}\int_{\Delta h_{\rm t}}^{+\infty}r\exp\left(-2\pi\lambda_{b}\int_{r}^{+\infty}
(\frac{1}{1+(\frac{\rho_{t}}{T})(\frac{x}{r})^{\alpha_{t}}})xdx-\pi\lambda_{b}(r^{2}-\Delta h_{\rm t}^2)\right)dr.
	\end{aligned}	
\end{equation}
\indent
\emph{Proof:} The result can be proved by a minor modification of Theorem 1 in \cite{ATAT}, where we consider the NOMA scheme for TU. $\hfill\square$
\subsection{Average Ergodic Rate of the Proposed CoMP-NOMA Scheme}
\indent
In this subsection, we derive the average ergodic rate to evaluate the network performance in terms of the spectral efficiency for the proposed CoMP-NOMA scheme in the cellular-connected UAV network. To be specific, we first define the average achievable rates for the typical Non-CoMP AU, CoMP AU, and TU as
\begin{equation}
	\label{Rate-expr}
	\mathcal{R}_{\rm u}^{\rm NC}\triangleq
\mathbb{E}_{\mathcal{B}}\left[\mathbb{E}_{\Upsilon_{\rm u}^{\rm NC}}\left[\log_{2}{\left(1+\Upsilon_{\rm u}^{\rm NC}\right)}\right]\right], \quad
	\mathcal{R}_{\rm u}^{\rm C}\triangleq
\mathbb{E}_{\mathcal{B}}\left[\mathbb{E}_{\Upsilon_{\rm u}^{\rm C}}\left[\log_{2}{\left(1+\Upsilon_{\rm u}^{\rm C}\right)}\right]\right], \quad
\end{equation}
\begin{equation}
	\mathcal{R}_{\rm t}\triangleq
\mathbb{E}_{\Upsilon_{\rm t}}\left[\log_{2}{\left(1+\Upsilon_{\rm t}\right)}\right].
\end{equation}
\indent
\emph{Theorem 3:} The average achievable rates for the typical Non-CoMP AU, CoMP AU, and TU are given by
\begin{equation}
	\label{Rate-expr-NC}
\mathcal{R}_{\rm u}^{\rm NC}
=\mathcal{A}_{\rm L_{0}}\mathcal{R}_{\rm u}^{\rm NC}(\mathcal{B}=\{b_{\rm L_0}\})+
\mathcal{A}_{\rm N_{0}}\mathcal{R}_{\rm u}^{\rm NC}(\mathcal{B}=\{b_{\rm N_0}\}),
\end{equation}
\begin{equation}
    \label{Rate-expr-C}
    \begin{aligned}
	\mathcal{R}_{\rm u}^{\rm C}&=
\mathcal{A}_{\rm L_{0},L_{1}}\mathcal{R}_{\rm u}^{\rm C}(\mathcal{B}=\{b_{\rm L_0}, b_{\rm L_1}\})+
\mathcal{A}_{\rm N_{0},N_{1}}\mathcal{R}_{\rm u}^{\rm C}(\mathcal{B}=\{b_{\rm N_0}, b_{\rm N_1}\})\\
&+\mathcal{A}_{\rm L_{0},N_{0}}\mathcal{R}_{\rm u}^{\rm C}(\mathcal{B}=\{b_{\rm L_0}, b_{\rm N_0}\})+
\mathcal{A}_{\rm N_{0},L_{0}}\mathcal{R}_{\rm u}^{\rm C}(\mathcal{B}=\{b_{\rm N_0}, b_{\rm L_0}\}),
    \end{aligned}
\end{equation}
\begin{equation}
    \label{Rate-expr-tu}
  \mathcal{R}_{\rm t}=2\pi\lambda_{b}\int_{0}^{+\infty}\int_{\Delta h_{\rm t}}^{+\infty}r\exp\left(-2\pi\lambda_{b}\int_{r}^{+\infty}
(\frac{1}{1+(\frac{\rho_{t}}{T})(\frac{x}{r})^{\alpha_{t}}})xdx-\pi\lambda_{b}(r^{2}-\Delta h_{\rm t}^2)\right)drd\tau,
\end{equation}
where the association probabilities are given by Lemma 3, and the conditional average achievable rates are given by
\begin{equation}\label{R-expr-NC}
\mathcal{R}_{\rm u}^{\rm NC}(\mathcal{B}=\{b_{\rm L_0}\})= \int_{0}^{+\infty}\mathbb P_{{\rm L_{0}}}(2^{\tau}-1)d\tau,\quad
\mathcal{R}_{\rm u}^{\rm NC}(\mathcal{B}=\{b_{\rm N_0}\})= \int_{0}^{+\infty}\mathbb P_{{\rm N_{0}}}(2^{\tau}-1)d\tau,
\end{equation}
\begin{equation}\label{R-expr-C}
\begin{aligned}
&\mathcal{R}_{\rm u}^{\rm C}(\mathcal{B}=\{b_{\rm L_0}, b_{\rm L_1}\})=\int_{0}^{+\infty}\mathbb P_{{\rm L_{0},L_{1}}}(2^{\tau}-1)d\tau,\quad
\mathcal{R}_{\rm u}^{\rm C}(\mathcal{B}=\{b_{\rm N_0}, b_{\rm N_1}\})=\int_{0}^{+\infty}\mathbb P_{{\rm N_{0},N_{1}}}(2^{\tau}-1)d\tau,\quad\\
&\mathcal{R}_{\rm u}^{\rm C}(\mathcal{B}=\{b_{\rm L_0}, b_{\rm N_0}\})=\int_{0}^{+\infty}\mathbb P_{{\rm L_{0},N_{0}}}(2^{\tau}-1)d\tau,\quad
\mathcal{R}_{\rm u}^{\rm C}(\mathcal{B}=\{b_{\rm N_0}, b_{\rm L_0}\})=\int_{0}^{+\infty}\mathbb P_{{\rm N_{0},L_{0}}}(2^{\tau}-1)d\tau,
\end{aligned}
\end{equation}
with the conditional coverage probabilities within (\ref{R-expr-NC}) and (\ref{R-expr-C}) being given by Theorem 1.\\
\indent
\emph{Proof.} The results can be proved by substituting $T \triangleq 2^{\tau}-1$ into the conditional coverage probabilities given in Theorem 2, and then integrating over the variable $\tau$. We omit the proof here due to space limitation. $\hfill\square$\\
\indent
According to the law of total probability, the average ergodic rate of the proposed CoMP-NOMA scheme is given by
\begin{equation}
\label{R-tot}
	\mathcal{R} = \mathcal{R}_{\rm u}^{\rm NC} +\mathcal{R}_{\rm u}^{\rm C} + \mathcal{R}_{\rm t}.
\end{equation}
By substituting $\mathcal{R}_{\rm u}^{\rm NC}$, $\mathcal{R}_{\rm u}^{\rm C}$ and $\mathcal{R}_{\rm t}$ given in (\ref{Rate-expr-NC}), (\ref{Rate-expr-C}), and (\ref{Rate-expr-tu}) into (\ref{R-tot}), we can derive the average ergodic rate.
 %  仿真与理论验证
\section{SIMULATIONS AND NUMERICAL RESULTS}
\indent
In this section, we first verify the validity of the proposed framework by means of simulations, and then show the effectiveness of the proposed CoMP-NOMA scheme in terms of coverage probability and average ergodic rate. To show the superiority of the proposed CoMP-NOMA scheme, we compare with three benchmark schemes, namely, CoMP-OMA scheme, NOMA-Only scheme, and OMA-Only scheme. To make a fair comparison, for the OMA in both CoMP-OMA scheme and OMA-Only scheme, the tagged BS allocates half unit of resource, and transmits with half of its power budget to both the typical AU and the typical TU, respectively \cite{DUNO}. While for the NOMA-Only scheme, all AUs are served as non-COMP AUs.
\begin{table} 
	\begin{center}
		\caption{Numerical and simulation parameters}
		\label{tab2}	
		\begin{tabular}{ |c|c|c| }
			\hline
			Parameter & Description & Value  \\ \cline{1-3}
			$ \alpha_{\rm L},\alpha_{\rm N}, \alpha_{\rm t} $ & Pathloss exponents for LoS, NLoS and TU links & 2.6, 3, 3  \\
			$ A_{\rm L},A_{\rm N},A_{\rm t} $ & Pathloss constants for LoS, NLoS, and TU links & -35 dB, -40 dB, -28.4 dB  \\
			$ m_{\rm L},m_{\rm N} $ & Nakagami-$m$ parameters for LoS and NLoS links & 3, 1 \\
			$P_{\rm t} $ & Transmit power of BS & 26 dB \\
			$\lambda_{\rm b} $ & Density of BSs & $ 10\ {\rm km}^{-2} $ \\
			$\rho_{\rm u}, \rho_{\rm t} $ & Power control coefficients for AU and TU & $ 0.9, 0.1 $ \\
			$\theta $ & Cooperation threshold & $4 \ {\rm dB}$ \\
			$h_{\rm u},h_{\rm b}, h_{\rm t}$ & Altitudes of AU, BS and TU & 75 m, 19 m, 1.5 m \\
			$B$, $C$ & Air-to-Ground channel parameters & 9.61, 0.16 \\
			\hline
		\end{tabular}
	\end{center}
\end{table}
\subsection{Analytical Framework Validation}
In this work, we employ the Gamma approximation and the Cauchy-Schwarz’s inequality in the analytical computation of coverage probability in Theorem 2. To verify the feasibility of these approximations, we consider a horizontal area of 10000 $\times$ 10000 ${\rm m}^2$ with $10^5$ iterations in the simulation. The horizontal locations of BSs, AUs, and TUs are a realization of three independent PPPs of densities $\lambda_{\rm b}$, $\lambda_{\rm u}$ and $\lambda_{\rm t}$, with $\lambda_{\rm t}$=10$\lambda_{\rm u}$=100$\lambda_{\rm b}$. Unless stated otherwise, we use the simulation parameters as listed in Table II.\\
\begin{figure}[t]
	%是可选项 h表示的是here在这里插入，t表示的是在页面的顶部插入
	\centering
	\includegraphics[scale=0.55]{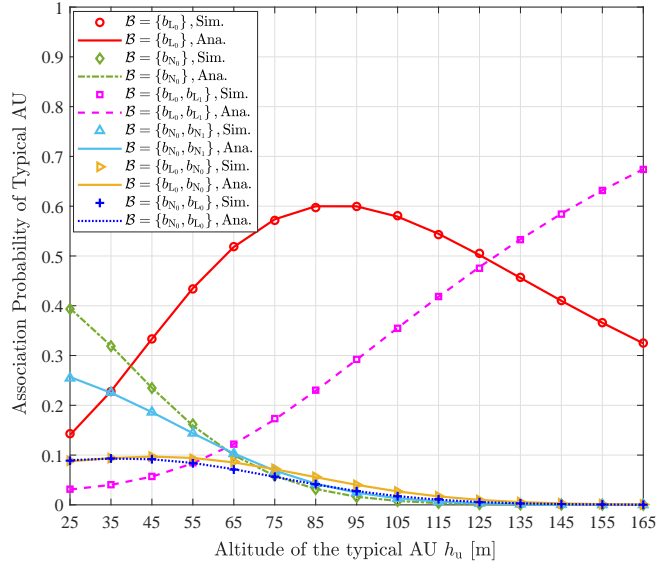}
	\caption{ Association probabilities of the typical AU as a function of AU's altitude for different association cases.}
	\label{fig-Asso-height}
\end{figure}
\indent
In Fig. \ref{fig-Asso-height}, we depict the association probabilities of AU as a function of AU's altitude for different association cases. It can be seen that the analytical results perfectly match the simulation results, which validates the accuracy of the obtained analysis in Lemma 3. We observe that $ \mathcal{A}_{{\rm L}_{0}} $ exhibits the concave behavior as a function of AU's altitude $ h_u $, and $ \mathcal{A}_{{\rm L}_{0},{\rm L}_{1}} $ grows with the increasing $ h_{\rm u} $, while the probabilities of the other four association cases $   \mathcal{A}_{{\rm N}_{0}} $, $\mathcal{A}_{{\rm N}_{0},{\rm N}_{1}}$, $\mathcal{A}_{{\rm L}_{0},{\rm N}_{1}}$, and $  \mathcal{A}_{{\rm N}_{0},{\rm L}_{1}} $ decline with the increasing $ h_{\rm u} $ and reduce to zero when $ h_{\rm u} $ is greater than 120 m. This is because with the increase of $ h_{\rm u} $, the LoS probability of AU grows which enlarges the probability to associate with the LoS BS. However, as $ h_{\rm u} $ further grows, $ \mathcal{A}_{{\rm L}_{0}} $ decreases while $ \mathcal{A}_{{\rm L}_{0},{\rm L}_{1}} $ increases. This is because the ratio of RSS from the nearest LoS BS to that from the dominant interfering BS declines, enlarging the cooperative probability. What's more, the gradually growing $  h_{\rm u} $ results in the higher LoS probability, leading to the higher $ \mathcal{A}_{{\rm L}_{0},{\rm L}_{1}} $.\\
\begin{figure} [t]
	\subfloat[\label{a}]{
		\includegraphics[scale=0.46]{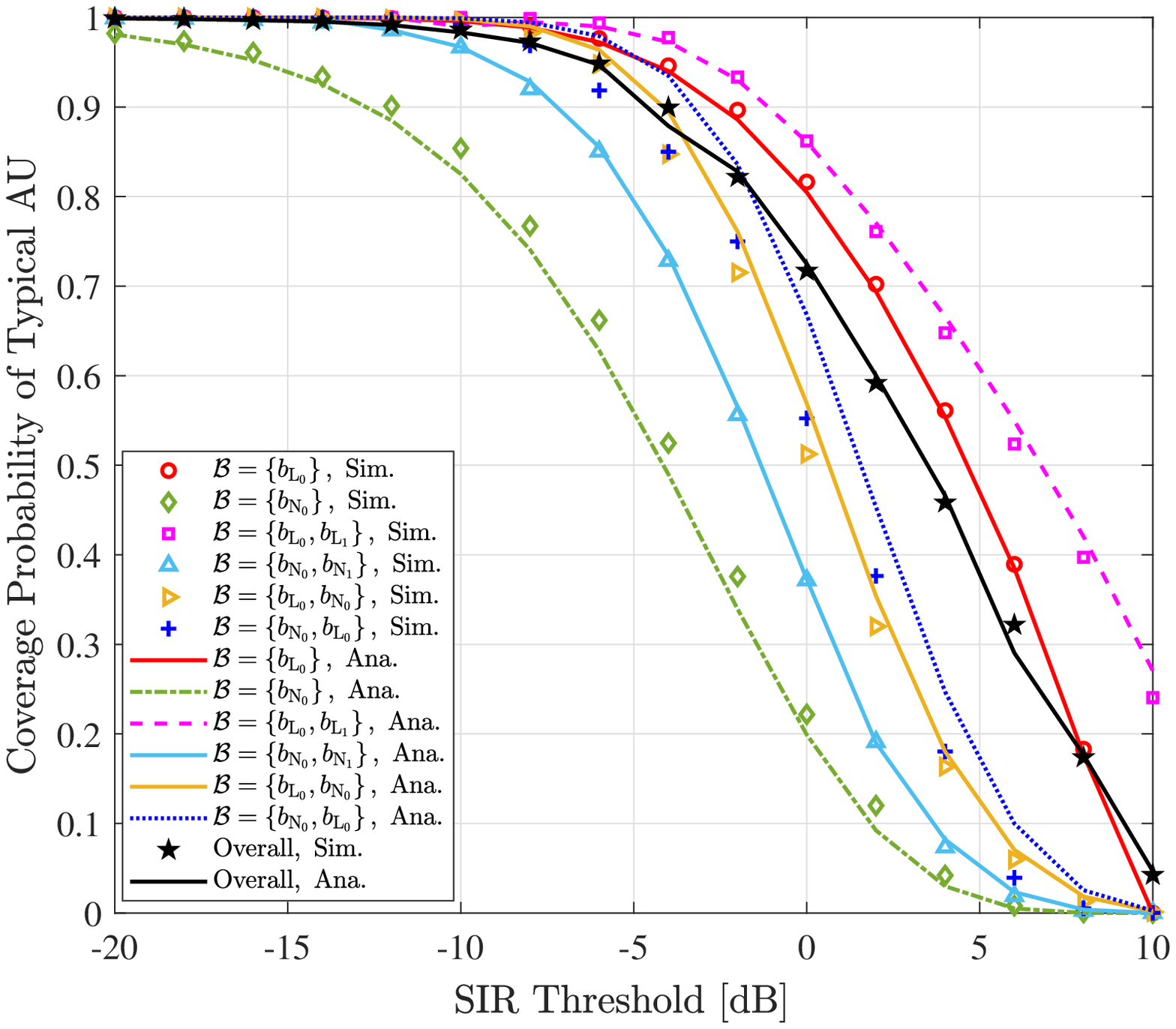}}
    \hspace{0.01\linewidth}
	\subfloat[\label{b}]{
		\includegraphics[scale=0.585]{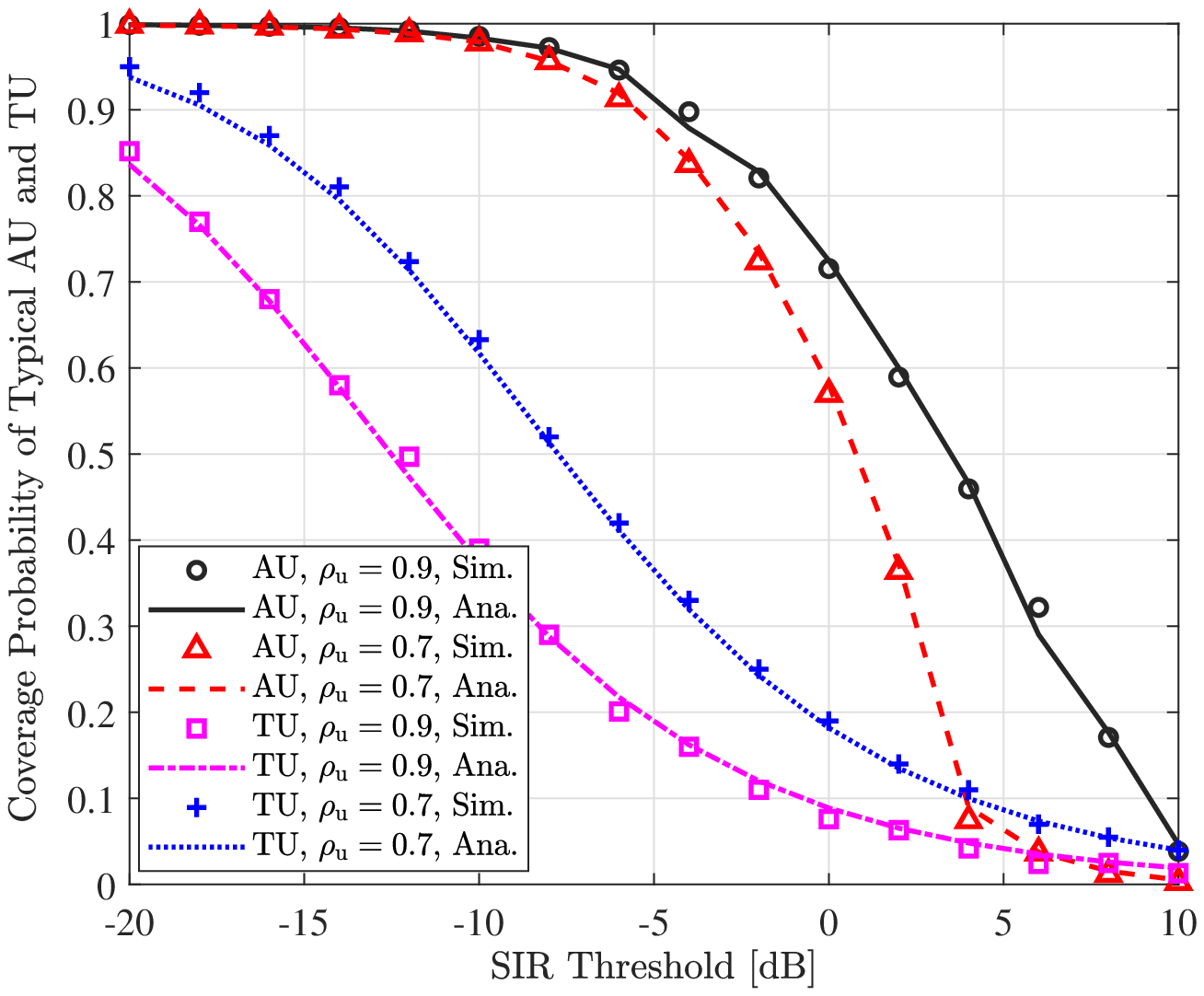} }
	\caption{(a) Conditional and overall coverage probabilities of the typical AU as a function of SIR threshold, (b) coverage probability of the typical AU and TU as a function of SIR threshold for different power control coefficient $\rho_{\rm u}$.}
    \label{fig-cov-AU-TU-SIR}
\end{figure}
\begin{comment}
    \begin{figure}[t]
	%是可选项 h表示的是here在这里插入，t表示的是在页面的顶部插入
	\centering
	\includegraphics[scale=0.5]{six_cov_AU_final_new.eps}
	\caption{(a) Conditional and overall coverage probabilities of the typical AU as a function of SIR threshold.}
	\label{fig-cov-AU-SIR}
\end{figure}
\end{comment}
\indent
In Fig. \ref{fig-cov-AU-TU-SIR}, we verify the accuracy of coverage probabilities of the typical AU and TU by varying SIR threshold and power control coefficient. Fig. \ref{fig-cov-AU-TU-SIR}(a) shows a good match between the simulations and the theoretical analysis for most of cases except for the cases $\mathcal{B}=\{b_{\rm N_0},b_{\rm L_0}\}$ and $\mathcal{B}=\{b_{\rm L_0},b_{\rm N_0}\}$. The gap is due to the use of Gamma approximation and Cauchy-Schwarz’s inequality in the computation of conditional coverage probability. However, since the occurrence probability of these two cases is small, the gap has little impact on the overall coverage probability. Associating with two LoS BSs (only one NLoS BS) achieves the largest (lowest) coverage for the AU. What's more, the only one LoS BS association case is the second largest, higher than the cooperation case with one LoS BS and one NLoS BS. This is because the only one LoS BS association case means that the largest RSS is sufficiently large, while the cooperation case means that the first two largest RSSs are comparable. 
%Note that the above relation is dependent on the cooperation threshold. 
Fig. \ref{fig-cov-AU-TU-SIR}(b) shows a good match between the simulations and analytical results, which verifies the correctness of the coverage probabilities for both the typical AU and TU. What's more, we observe that increasing $\rho_{\rm u}$ is beneficial to AU's coverage probability while aggravating the TU's coverage probability. This is due to the fact that allocating more transmit power to the typical AU enhances its received SIR.
\begin{comment}
 \begin{figure}[t]
	%是可选项 h表示的是here在这里插入，t表示的是在页面的顶部插入
	\centering
	\includegraphics[scale=0.65]{cov_1e5-AU-TU.eps}
	\caption{Coverage probability of the typical AU and TU as a function of SIR threshold for different power control coefficient $\rho_{\rm u}$ with $\lambda_{\rm b}=10^{-5}\ {\rm m}^{-2}$.}
	\label{fig-cov-AU-TU-SIR}
\end{figure}\\   
\end{comment}

\subsection{Coverage Probability Evaluation}
\begin{figure}[t]
	%是可选项 h表示的是here在这里插入，t表示的是在页面的顶部插入
	\centering
	\includegraphics[scale=0.7]{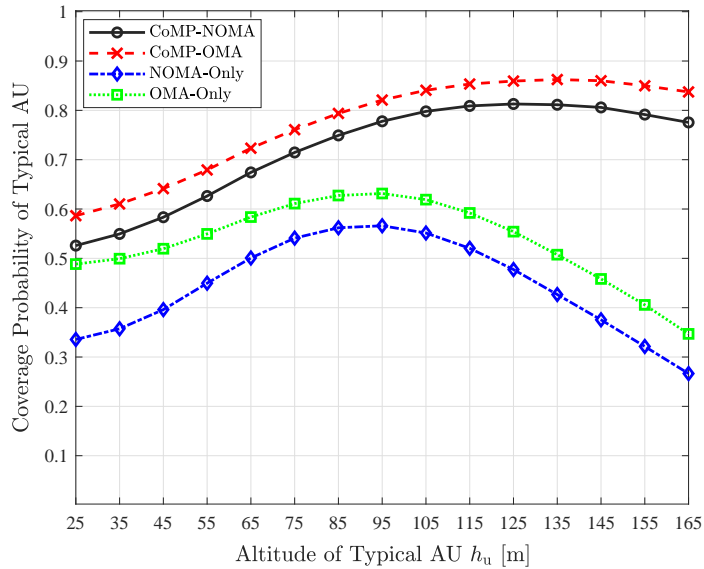}
	\caption{Comparison of the typical AU's coverage probability for different schemes as a function of AU's altitude.}
	\label{fig-cov-AU-comp-height}
\end{figure}
\indent
In Fig. \ref{fig-cov-AU-comp-height}, we evaluate the coverage probability of AU as a function of AU's altitude under different schemes. We observe that the coverage probability achieved by the proposed CoMP-NOMA scheme is higher than that achieved by NOMA-Only scheme and OMA-Only scheme, while a little lower than that achieved by the CoMP-OMA scheme. This is due to the enhanced SIR achieved by CoMP, and the inter-user interference caused by NOMA. What's more, the coverage probability first grows and then declines as a function of the AU's altitude. The augmented coverage probability in the initial stage is due to the growing LoS probability of the A2G link and thus the received SIR, while the decreasing coverage probability in the later stage can be explained by the fact that the incremental path loss overweighs the gain from the increasing LoS probability. Therefore, the proposed framework allows to derive the optimal AU's altitude to achieve the largest coverage probability for AU. \\
\begin{figure} [t]
	\subfloat[\label{a}]{
		\includegraphics[scale=0.5]{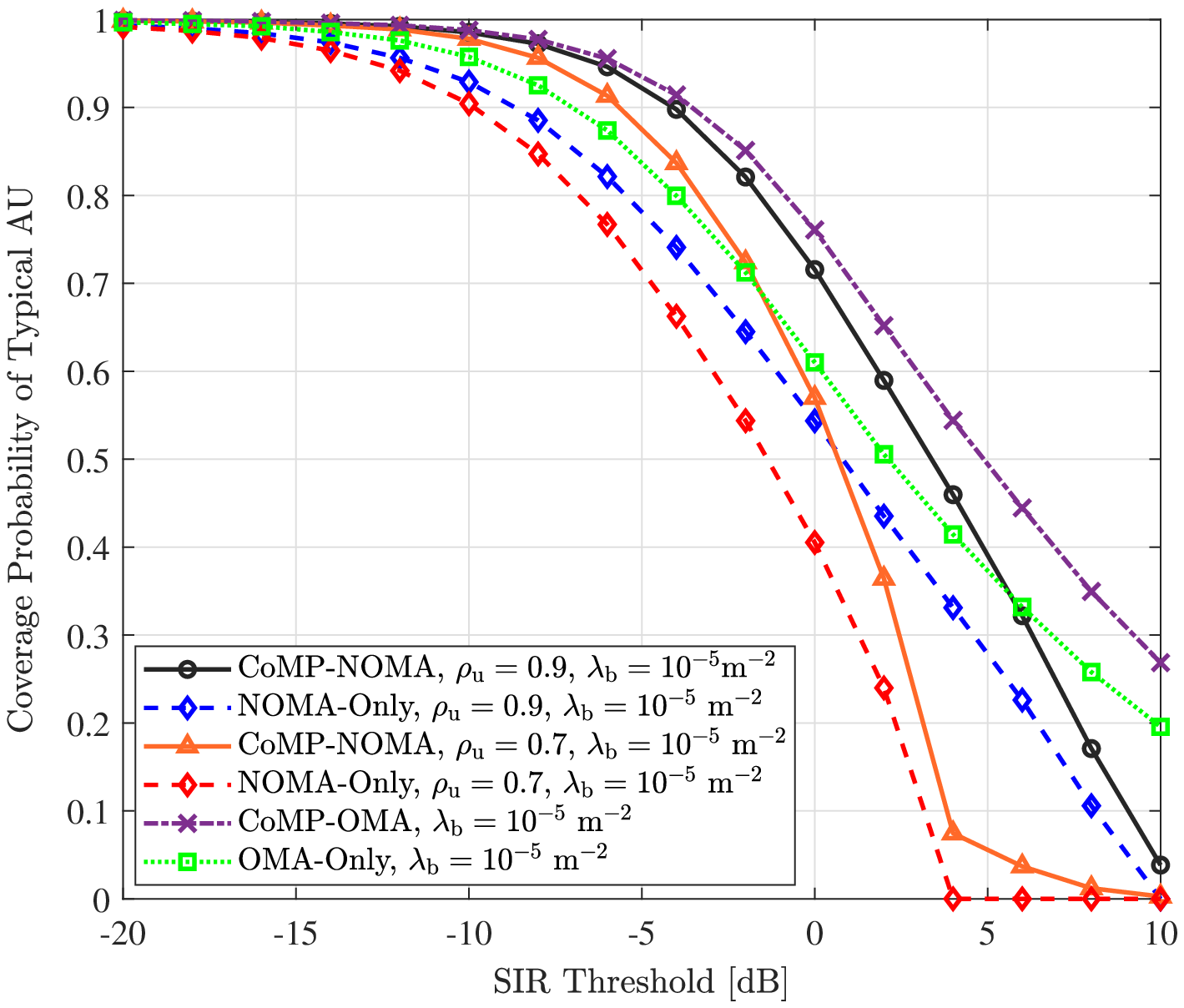}}
    \hspace{0.01\linewidth}
	\subfloat[\label{b}]{
		\includegraphics[scale=0.5]{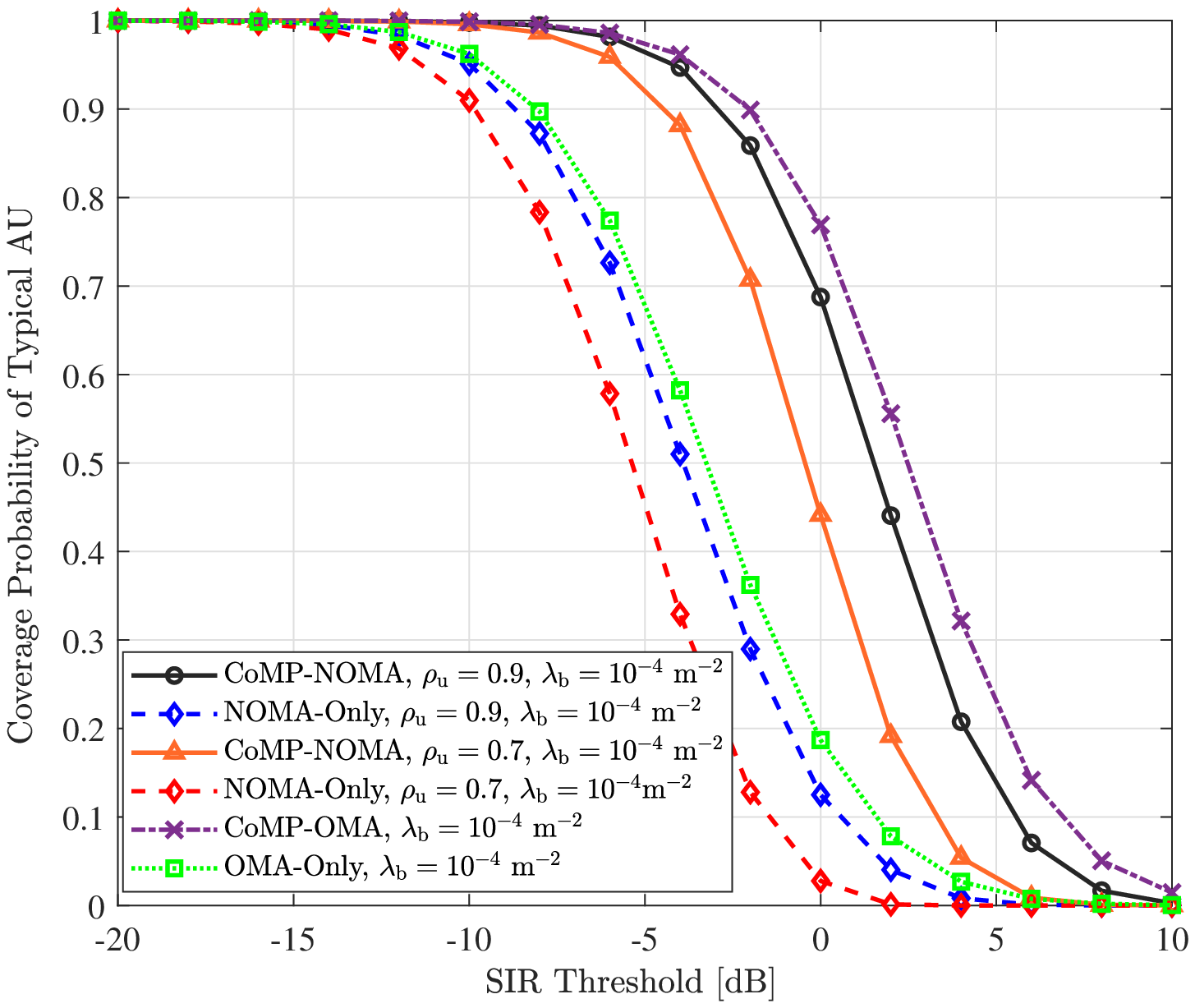} }
	\caption{Comparison of the typical AU's coverage probability for different schemes as a function of SIR threshold, where (a) is for $\lambda_{\rm b}=10^{-5}\ {\rm m}^{-2}$, and (b) is for $\lambda_{b}=10^{-4}\ {\rm m}^{-2}$.}
    \label{fig-cov-AU-comp-SIR}
\end{figure}
\indent
In Fig. \ref{fig-cov-AU-comp-SIR}, we compare the typical AU's coverage probability for different schemes as a function of SIR threshold with different power control coefficient $\rho_{\rm u}$ and BS density $\lambda_{\rm b}$. For both Fig. \ref{fig-cov-AU-comp-SIR}(a) and Fig. \ref{fig-cov-AU-comp-SIR}(b), we observe that the COMP-OMA scheme achieves the highest coverage probability, while the NOMA-Only scheme with $\rho_{\rm u}$=0.7 achieves the lowest coverage probability. Allocating more power budget to AU in both CoMP-NOMA scheme and NOMA-Only scheme is beneficial to AU's coverage probability. For a smaller BS density $\lambda_{\rm b}=10^{-5}\ {\rm m}^{-2}$, Fig. \ref{fig-cov-AU-comp-SIR}(a) shows that $\rho_{\rm u}$ has larger effect for a higher SIR threshold. That's why the typical AU's coverage probability achieved by COMP-NOMA scheme with $\rho_{\rm u}$=0.7 is lower than that achieved by NOMA-Only scheme with $\rho_{\rm u}$=0.9 and OMA-Only scheme when the SIR threshold increases to a certain extent. Yet, for a larger BS density $\lambda_{\rm b}=10^{-4}\ {\rm m}^{-2}$, we observe that there is no crossing between curves. This is because the network is interference-limited in the dense network scenario, which highlights the benefit of BS cooperation in reducing the dominated interferer.

\begin{comment}
 \begin{figure}[h]
	%是可选项 h表示的是here在这里插入，t表示的是在页面的顶部插入
	\centering
	\includegraphics[scale=0.6]{Rate_height_final-0405.eps}
	\caption{Average achievable rate for CoMP AU, non-CoMP AU, Overall AU, and TU as a function of AU's altitude.}
	\label{fig-rate-AU-TU-height}
\end{figure}
\begin{figure}[t]
	%是可选项 h表示的是here在这里插入，t表示的是在页面的顶部插入
	\centering
	\includegraphics[scale=0.6]{Rate_theta_final-0405.eps}
	\caption{Average achievable rate for CoMP AU, non-CoMP AU, and TU as a function of cooperation threshold $\theta$.}
	\label{Rate-AU-TU-theta}
\end{figure}   
\end{comment}
\begin{figure} [t]
	\subfloat[\label{a}]{
		\includegraphics[scale=0.5]{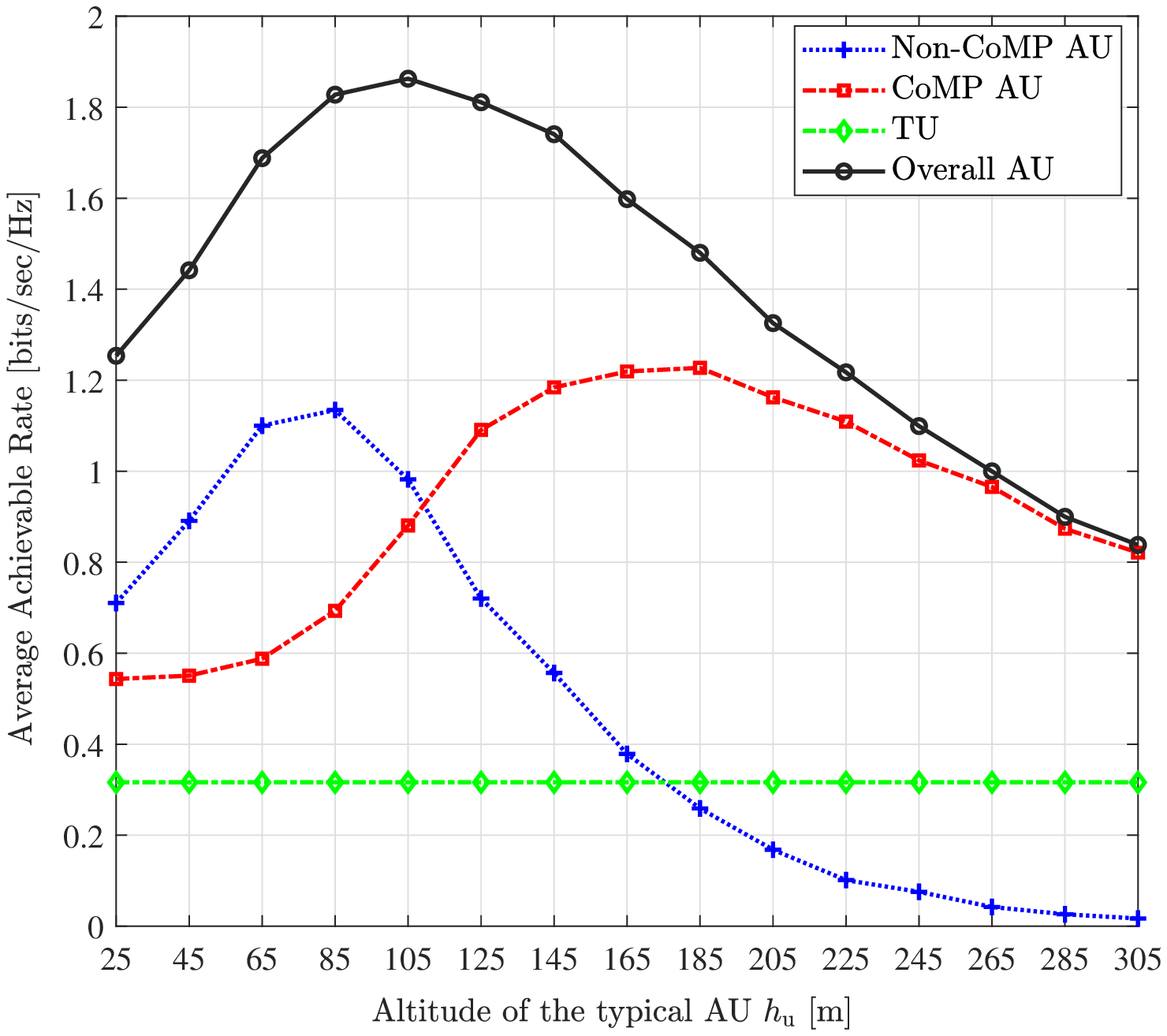}}
    \hspace{0.01\linewidth}
	\subfloat[\label{b}]{
		\includegraphics[scale=0.5]{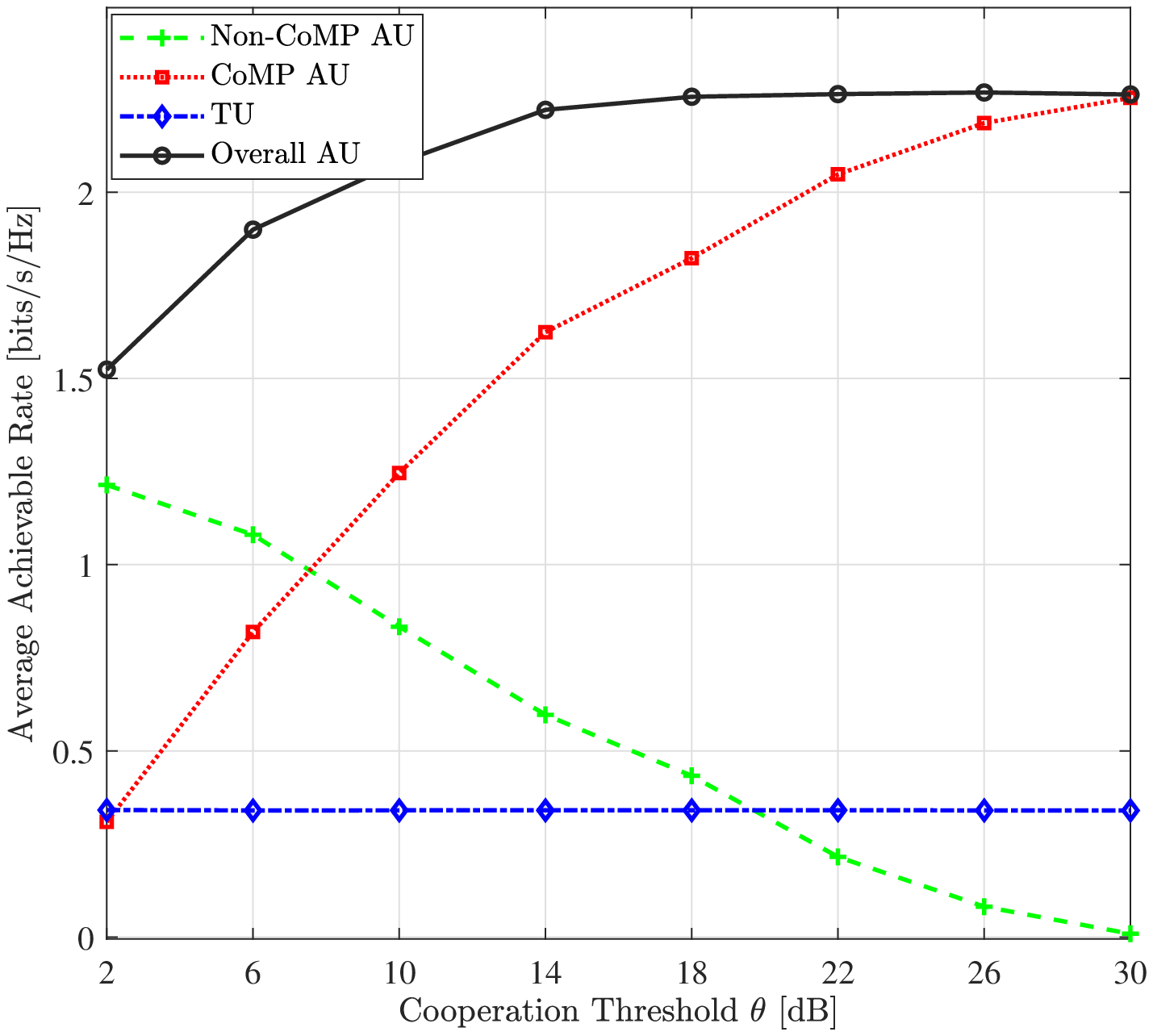} }
	\caption{Average achievable rate for CoMP AU, non-CoMP AU, Overall AU, and TU by varying (a) AU's altitude, and (b) cooperation threshold $\theta$.}
    \label{fig-rate-AU-TU}
\end{figure}
\subsection{Average Ergodic Rate Evaluation}

In Fig. \ref{fig-rate-AU-TU}, we depict the average achievable rates of the typical Non-CoMP AU ($\mathcal{R}_{\rm u}^{\rm NC}$), CoMP AU ($\mathcal{R}_{\rm u}^{\rm C}$), Overall AU ($\mathcal{R}_{\rm u}^{\rm NC}$+$\mathcal{R}_{\rm u}^{\rm C}$), and typical TU ($\mathcal{R}_{\rm t}$) given in Theorem 3 as a function of the typical AU's altitude $h_{\rm u}$ and the cooperation threshold $\theta$, respectively. In Fig. \ref{fig-rate-AU-TU}(a), as $h_{\rm u}$ grows, we observe that $\mathcal{R}_{\rm u}^{\rm NC}$ and $\mathcal{R}_{\rm u}^{\rm C}$ increase first and then decrease, leading to the same trend of  $\mathcal{R}_{\rm u}^{\rm NC}$+$\mathcal{R}_{\rm u}^{\rm C}$, while $\mathcal{R}_{\rm t}$ keeps unchanged. The variation of $\mathcal{R}_{\rm u}^{\rm NC}$ and $\mathcal{R}_{\rm u}^{\rm C}$ is due to the tradeoff between the incremental LoS probability for A2G link and the enlarging path loss. When $h_{\rm u}$ achieves 305 m, we observe that only CoMP AUs contribute to the rate. This can be explained by the fact that as $h_{\rm u}$ rises, the first two largest average RSSs become more comparable, increasing the probability of being a CoMP AU. The unchanged curve for $\mathcal{R}_{\rm t}$ can be explained by the assumption of perfect SIC conducted by TU, which eliminates the inter-user interference in NOMA. In Fig. \ref{fig-rate-AU-TU}(b), as $\theta$ increases, we observe that $\mathcal{R}_{\rm u}^{\rm NC}$ decreases and $\mathcal{R}_{\rm u}^{\rm C}$ increases, leading to an increase in $\mathcal{R}_{\rm u}^{\rm NC}$+$\mathcal{R}_{\rm u}^{\rm C}$. This is due to the fact that a higher $\theta$ enlarges the probability of being a CoMP AU. When $\theta$ grows to a certain value, e.g., 18 dB in this example, nearly all AUs are CoMP AUs, leading to a convergence of the overall rate. 

\begin{comment}
  \begin{figure}[t]
	%是可选项 h表示的是here在这里插入，t表示的是在页面的顶部插入
	\centering
	\includegraphics[scale=0.6]{Ergodic_rate_height_Compare_final.eps}
	\caption{Average achievable rate for different schemes as a function of AU's altitude.}
	\label{Ergodic-rate-comp-height}
\end{figure}
\begin{figure}[t]
	%是可选项 h表示的是here在这里插入，t表示的是在页面的顶部插入
	\centering
	\includegraphics[scale=0.6]{Ergodic_Rate_theta_Compare_final.eps}
	\caption{Comparison of average ergodic rate for different schemes as a function of cooperation threshold $\theta$.}
	\label{Ergodic-rate-comp-theta}
\end{figure}\\  
\end{comment}
\begin{figure} [t]
	\subfloat[\label{a}]{
		\includegraphics[scale=0.5]{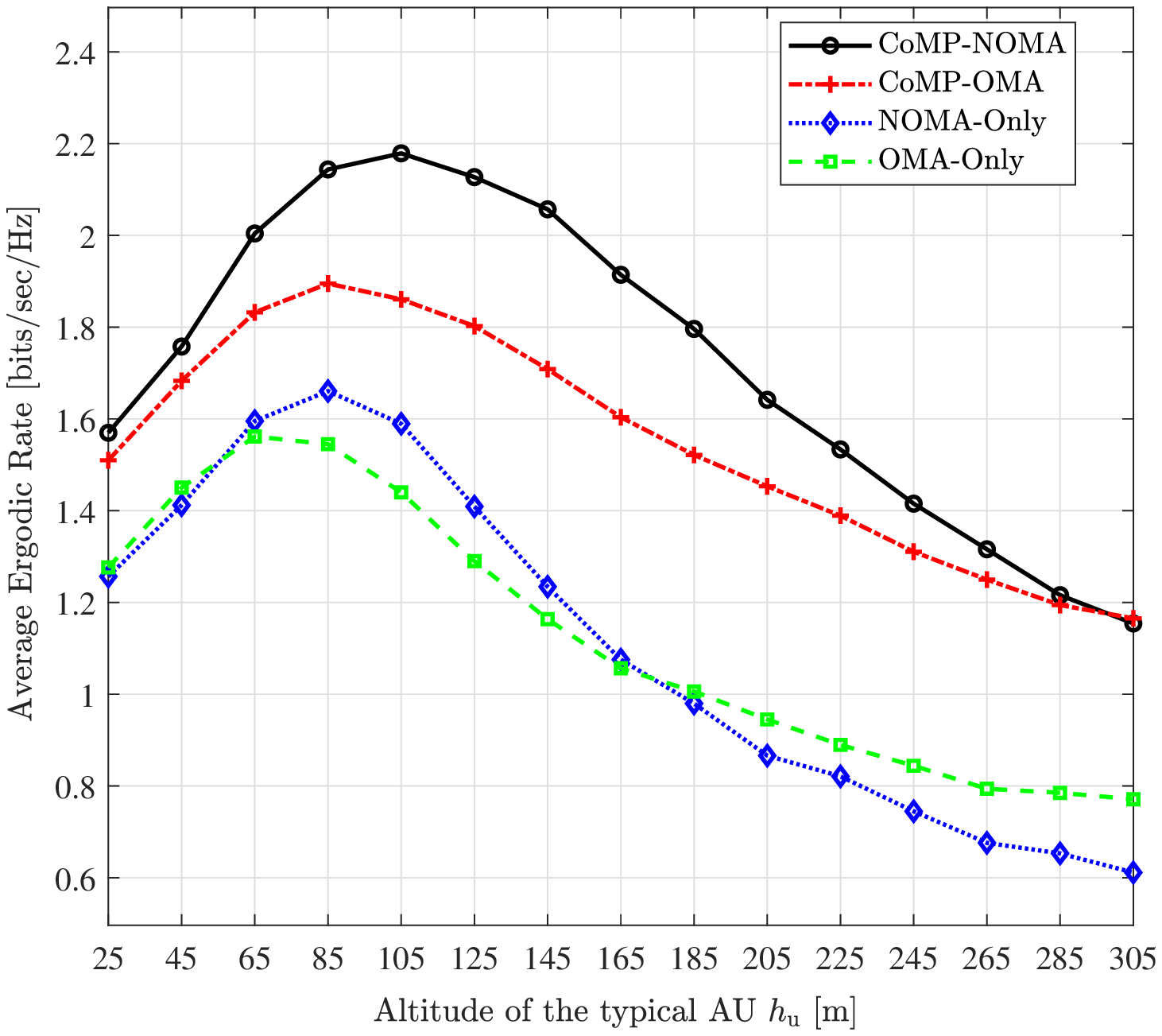}}
    \hspace{0.01\linewidth}
	\subfloat[\label{b}]{
		\includegraphics[scale=0.5]{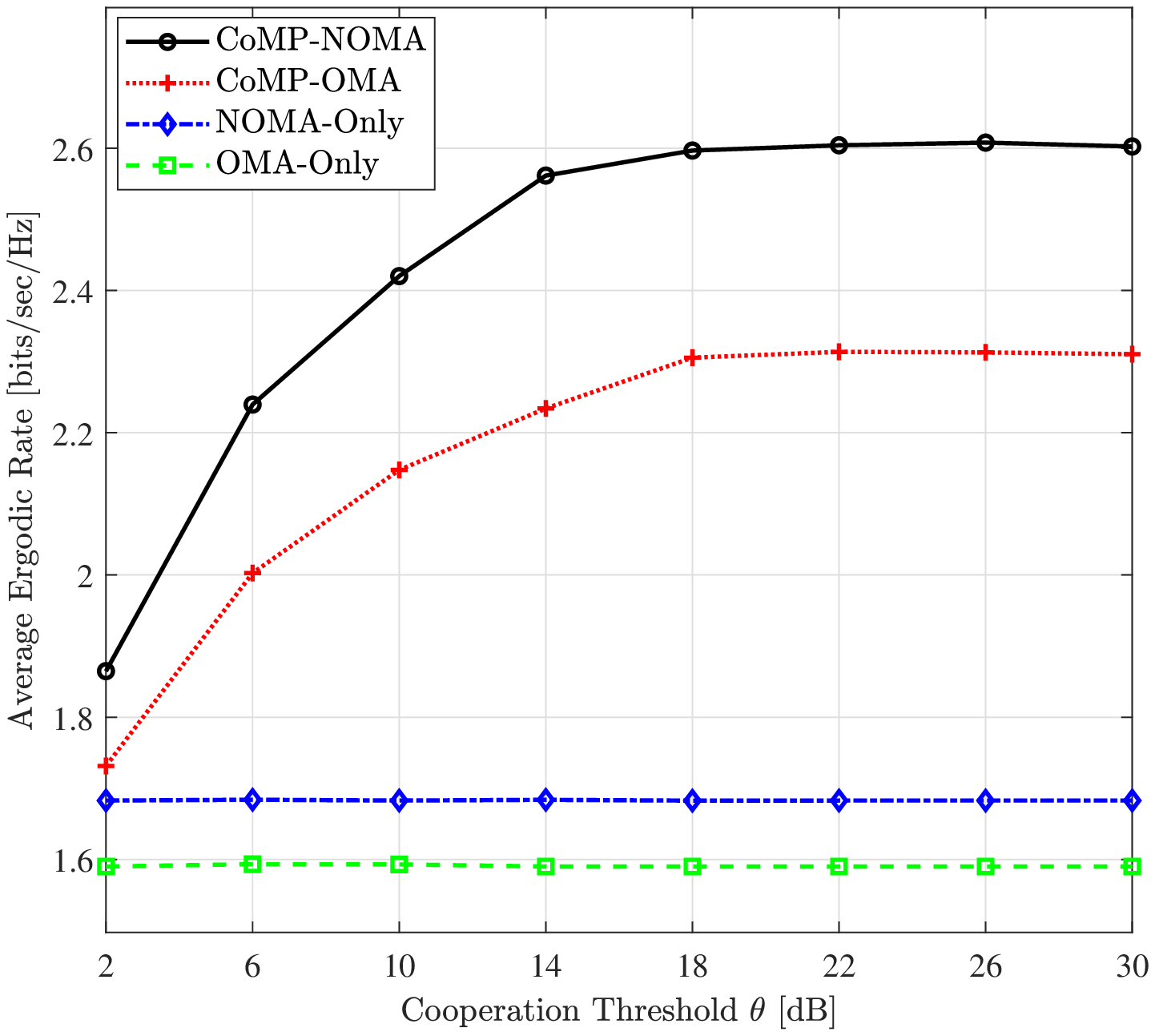}}
	\caption{Average achievable rate for different schemes by varying (a) AU's altitude, and (b) cooperation threshold $\theta$.}
    \label{fig-Ergodic-rate-comp}
\end{figure}
In Fig. \ref{fig-Ergodic-rate-comp}, we demonstrate the effectiveness of the proposed CoMP-NOMA scheme in terms of the average ergodic rate $\mathcal{R}$ in (\ref{R-tot}) by varying the altitude of the typical AU $h_{\rm u}$ and cooperation threshold $\theta$, respectively. In Fig. \ref{fig-Ergodic-rate-comp}(a), we observe that the ergodic rate achieved by the proposed CoMP-NOMA scheme is the largest, which is more superior for an appropriate $h_{\rm u}$. What's more, we observe that the OMA-Only scheme may achieve the same or even higher ergodic rate than the NOMA-Only scheme when $h_{\rm u}$ is small or larger than a certain value. This can be explained by the fact that the inter-user interference in NOMA has a great impact on the typical AU, which is even obvious for a higher altitude. It shows that an appropriate setting of AU's altitude is beneficial to maximize the gain achieved by the proposed COMP-NOMA scheme. Fig. \ref{fig-Ergodic-rate-comp}(b) shows that $\mathcal{R}$ achieved by CoMP-NOMA scheme and CoMP-OMA scheme both grow with $\theta$, and converge when $\theta$ increases to a certain value, while $\mathcal{R}$ keeps unchanged for both NOMA-Only scheme and OMA-Only scheme. This is because when $\theta$ grows, the probability of the typical AU being a CoMP AU increases, harvesting more gain from the BS cooperation. When the probability of being a CoMP AU approaches to 1, the average ergodic rate converges to a constant value. Since there is no BS cooperation in NOMA-Only scheme and OMA-Only scheme, $\theta$ has no effect on $\mathcal{R}$. Although increasing the CoMP probability is beneficial to boost the average ergodic rate of the system, it also enlarges the cooperation overhead of the system. Our proposed analytical framework can be used to determine an appropriate cooperation threshold to maximize the ergodic rate while maintaining a relatively lower cooperation overhead.  

\section{CONCLUSION}
In this work, we proposed an interference-aware CoMP-NOMA scheme for cellular-connected UAV networks by considering the harmonious coexist of AUs and TUs. In this scheme, we exploited the BS cooperation gain for the qualified AUs by considering the CoMP scheme, while NOMA scheme was employed to enable the nonorthogonal transmissions for AUs and TUs by leveraging SIC. We first designed the classification rule for AUs, and the rule of NOMA cluster formulation for AUs and TUs. We then proposed an analytical framework to evaluate the coverage probability and spectral efficiency of the proposed CoMP-NOMA scheme. The superiority of the proposed CoMP-NOMA scheme has been demonstrated by comparing with three benchmark schemes. Our results showed that the proposed scheme can  enhance the reliability of AUs through CoMP, and improve the spectral efficiency through NOMA as well. This work can
be extended by considering the design of NOMA pairing strategy to further enhance the network performance, and including the cooperation overhead evaluation to enrich the analytical framework. 

\appendix{}
\begin{comment}
\emph{A. Proof of Lemma 1}\\
\indent
We define the 3D distance between the typical Non-CoMP AU and the nearest NLoS BS within $\Phi^{\rm N}_{\rm B}$ as $R_{\rm N_{0}}$, of which the horizontal distance is expressed as $Z_{\rm N_{0}} =\sqrt{R_{\rm N_{0}}^{2}-(\Delta h_{\rm u})^{2}}$. Thus, the cumulative distribution function (CDF) of $R_{\rm N_{0}}$ is 
\begin{equation}\label{Lemma1}
		F_{R_{{\rm N}_{0}}}(r) = 1-\text{Pr}(Z_{{\rm N}_{0}} > \sqrt{{r}^{2}-(\Delta h_{\rm u})^{2}})\overset{\text{(a)}}{=} 1-\exp\bigg(-2\pi\lambda_{b}\int_{0}^{\sqrt{{r}^{2}-(\Delta h_{\rm u})^{2}}}zp^{{\rm N}}(z)dz\bigg),
\end{equation}
where (a) is due to the null probability of PPP. By definition, we finally derive $f_{R_{{\rm N}_{0}}}(r)\triangleq\dfrac{d}{dr} F_{R_{{\rm N}_{0}}}(r)$ in (\ref{fun13}). Following the same steps as of $f_{R_{{\rm N}_{0}}}(r)$, we can derive the PDF $f_{R_{{\rm L}_{0}}}(r)$ of distance between the typical Non-CoMP AU and the nearest LoS BS within $\Phi^{\rm L}_{\rm B}$ in (\ref{fun14}), which completes the proof.
\end{comment}

\emph{A. Proof of Lemma 2}\\
\indent
For the typical CoMP AU, there exists four cases for the cooperative BS set $\mathcal{B}$. We first consider the case $\mathcal{B}=\{b_{\rm N_{0}}, b_{\rm N_{1}}\}$, and define $R_{\rm N_{0}}$, and $R_{\rm N_{1}}$ as the 3D distances between the typical CoMP AU and the first two nearest NLoS BSs within $\Phi^{\rm N_{B}}$, respectively. Meanwhile, $f_{R_{\rm N_{0}},R_{\rm N_{1}}}(r_{\rm N_0},r_{\rm N_1})$ is defined as the joint PDF of the distances. By definition, we have
\begin{equation}\label{Lemma2-a}
f_{R_{{\rm N}_{0}},R_{{\rm N}_{1}}}(r_{{\rm N}_{0}},r_{{\rm N}_{1}}) = f_{R_{{\rm N}_{1}}|R_{\rm N_{0}}}(r_{{\rm N}_{1}} | r_{{\rm N}_{0}})f_{R_{{\rm N}_{0}}}(r_{{\rm N}_{0}}), \quad r_{\rm N_{1}} > r_{\rm N_{0}} \ge \Delta h_{\rm u},
\end{equation}
where the conditional PDF in the righthand is given by
\begin{equation}\label{Lemma2-b}
	 f_{R_{{\rm N}_{1}}|R_{\rm N_{0}}}\left(r_{\rm N_1} | r_{\rm N_0}\right) = 2\pi\lambda_{b}r_{\rm N_1}p^{{\rm N}}\left(r_{\rm N_1}\right)\exp\biggl(-2\pi\lambda_{b}\bigg(\int_{0}^{l(r_{\rm N_1})}zp^{{\rm L}}(z)dz- \int_{0}^{l(r_{\rm N_0})}zp^{{\rm L}}(z)dz\bigg)\biggl).
\end{equation}
By substituting (\ref{Lemma2-b}) into (\ref{Lemma2-a}), we complete the proof of $f_{R_{{\rm N}_{0}},R_{{\rm N}_{1}}}(r_{{\rm N}_{0}},r_{{\rm N}_{1}})$ in (\ref{fun15}). Following the same steps as of $f_{R_{{\rm N}_{0}},R_{{\rm N}_{1}}}(r_{{\rm N}_{0}},r_{{\rm N}_{1}})$, we can complete the proof of $f_{R_{{\rm L}_{0}},R_{{\rm L}_{1}}}(r_{{\rm L}_{0}},r_{{\rm L}_{1}})$, $f_{R_{{\rm N}_{0}},R_{{\rm L}_{0}}}(r_{{\rm N}_{0}},r_{{\rm L}_{0}})$, and $f_{R_{{\rm L}_{0}},R_{{\rm N}_{0}}}(r_{\rm L_0},r_{\rm N_0})$, given by (\ref{fun16}), (\ref{fun17}), and (\ref{fun18}), respectively. 
%Finally, we prove $f_{R_{{\rm L}_{0}},R_{{\rm N}_{0}}}(r_{{\rm L}_{0}},r_{{\rm N}_{0}})$. Based on the association strategy, if $r_{\rm L_{0}} \in [\Delta h_{u}, l_{\rm L\_N}]$ with $l_{\rm L\_N}\triangleq(\dfrac{\eta_{\rm L}}{\eta_{\rm N}})^{\frac{1}{\alpha_{\rm L}}}(\Delta h_{\rm u})^{\frac{\alpha_{\rm N}}{\alpha_{\rm L}}}$, the position of the closest NLoS BS is independent of the closest LoS BS, leading to the result $f_{R_{{\rm L}_{0}},R_{{\rm N}_{0}}}(r_{\rm L_0},r_{\rm N_0}) = f_{R_{{\rm N}_{0}}}(r_{\rm N_0})f_{R_{\rm L_{0}}}(r_{\rm L_{0}})$. Otherwise, if $r_{\rm L_{0}} \ge l_{\rm L\_N}$, following the same steps as of $f_{R_{{\rm N}_{0}},R_{{\rm N}_{1}}}(r_{{\rm N}_{0}},r_{{\rm N}_{1}})$, we complete the proof of $f_{R_{{\rm L}_{0}},R_{{\rm N}_{0}}}(r_{\rm L_0},r_{\rm N_0})$ in (\ref{fun18}).

\emph{B. Proof of Lemma 3}
\\
\indent
The probability of the typical Non-CoMP AU being associated with $\{ b_{\rm L_{0}}\}$ is given by
\begin{equation}
 \begin{aligned}
  \mathcal{A}_{\rm L_{0}} &= \text{Pr}\left [\mathcal{B}=\{b_{\rm L_{0}}\}\right] = \mathbb{E}_{R_{\rm L_{0}}}\biggl [\text{Pr}\bigg(\frac{\eta_{\rm L}R_{\rm L_{0}}^{-\alpha_{\rm L}}}{\eta_{\rm I_{n}}R_{\rm I_{n}}^{-\alpha_{\rm I_{n}}}}\ge \theta \bigg)\biggl]\\
  %&=\mathbb{E}_{R_{\rm L_{0}}}\left [\text{Pr}\left (\eta_{\rm L}R_{\rm L_{0}}^{-\alpha_{\rm L}} \ge \theta \max(\eta_{\rm L}R_{\rm L_{1}}^{-\alpha_{\rm L}} ,\eta_{\rm N}R_{\rm N_{0}}^{-\alpha_{\rm N}} )\right)\right ]\\
  &=\int_{\Delta h_{\rm u}}^{+\infty}\text{Pr}\left [R_{\rm L_{1}} \ge \theta^{\frac{1}{\alpha_{\rm L}}}R_{\rm L_{0}}\right ]
  \text{Pr}\left [R_{\rm N_{0}} \ge (\theta\frac{\eta_{\rm N}}{\eta_{\rm L}})^{\frac{1}{\alpha_{\rm N}}}R_{\rm L_{0}}^{\frac{\alpha_{\rm L}}{\alpha_{\rm N}}} \right ] f_{R_{\rm L_{0}}}(r_{\rm L_{0}})dr_{\rm L_{0}}    \\
  &\stackrel{\text{(a)}}{=}\int_{\Delta h_{\rm u}}^{+\infty}\int_{\theta^{\frac{1}{\alpha_{\rm L}}}r_{\rm L_{0}}}^{+\infty} f_{R_{{\rm L}_{1}}|R_{\rm L_{0}}}(r_{{\rm L}_{1}} | r_{{\rm L}_{0}})\exp\bigg(-2\pi\lambda_{b}\int_{0}^{l(\widetilde{d}_{\rm L\_N}(r_{\rm L_{0}}))}zp^{\rm N}(z)dz\bigg) f_{R_{\rm L_{0}}}(r_{\rm L_{0}})dr_{\rm L_{1}}dr_{\rm L_{0}}, 
 % &\stackrel{\text{(b)}}{=}\int_{\Delta h_{\rm u}}^{l_{\rm L\_N}} \int_{\theta^{\frac{1}{\alpha_{\rm L}}}r_{\rm L_{0}}}^{+\infty}f_{R_{\rm L_{0}},R_{\rm L_{1}}}(r_{\rm L_{0}},r_{\rm L_{1}})dr_{\rm L_{1}}dr_{\rm L_{0}}\\
 % &+\int_{l_{\rm L\_N}}^{+\infty}\int_{\theta^{\frac{1}{\alpha_{\rm L}}}r_{\rm L_{0}}}^{+\infty}f_{R_{\rm L_{0}},R_{\rm L_{1}}}(r_{\rm L_{0}},r_{\rm L_{1}})\exp\left(-2\pi\lambda_{b}\int_{0}^{l(\theta^{\frac{1}{\alpha_{\rm N}}}d_{\rm L\_N}(r_{\rm L_{0}}))}zp^{\rm N}(z)dz\right)dr_{\rm L_{1}}dr_{\rm L_{0}},
 \end{aligned}
\end{equation}
where (a) follows from the joint PDF and null probability of inhomogenous PPP $\Phi^{\rm N}_{\rm B}$, and we define $l(r)\triangleq \sqrt{r^2-(\Delta h_{\rm u})^2}$,
\begin{equation}
\label{d_rL0}
\widetilde{d}_{\rm L\_N}(r_{\rm L_{0}}) \triangleq \begin{cases}
\Delta h_{\rm u}, &{\text{if}} \quad \Delta h_{\rm u}\le r_{\rm L_{0}} < l_{\rm L\_{N}},\\
\theta^{\frac{1}{\alpha_{\rm N}}}d_{\rm L\_N}(r_{\rm L_{0}}), &{\text{if}} \quad r_{\rm L_{0}} \ge l_{\rm L\_{N}},
\end{cases}
\end{equation}
 with $d_{\rm L\_N}(r_{\rm L_{0}})\triangleq (\dfrac{\eta_{\rm N}}{\eta_{\rm L}})^{\frac{1}{\alpha_{\rm N}}}r_{\rm L_{0}}^{\frac{\alpha_{\rm L}}{\alpha_{\rm N}}}$, $l_{\rm L\_N}\triangleq(\dfrac{\eta_{\rm L}}{\eta_{\rm N}})^{\frac{1}{\alpha_{\rm L}}}(\Delta h_{\rm u})^{\frac{\alpha_{\rm N}}{\alpha_{\rm L}}}$. Simplifying the expression by considering the value range of $r_{\rm L_{0}}$, we complete the proof of $\mathcal{A}_{\rm L_{0}}$ in (\ref{fun19}). Following the similar steps, we derive the probability that the typical Non-CoMP AU is associated with $\{ b_{\rm N_{0}}\}$ in (\ref{fun20}).\\
\indent
Then, we derive the association probability when the typical AU is associated with $\{b_{\rm L_{0}},b_{\rm L_{1}}\}$, which is given by
\begin{equation}
   \begin{aligned}
\mathcal{A}_{\rm L_{0},L_{1}}
&=\text{Pr}\left[\mathcal{B}=\{b_{\rm L_{0}}, b_{\rm L_{1}}\}\right] 
%=\mathbb{E}_{R_{\rm L_{0}},R_{\rm L_{1}}}\left
%[\text{Pr}\left [\theta R_{{\rm L_{1}}}^{-\alpha_{\rm L}}\ge R_{{\rm L_{0}}}^{-\alpha_{\rm L}},\eta_{\rm L}R_{{\rm L_{1}}}^{-\alpha_{\rm L}}\ge \eta_{\rm N}R_{{\rm N_{0}}}^{-\alpha_{\rm N}}\right]\right]\\
&=\mathbb{E}_{R_{\rm L_{0}},R_{\rm L_{1}}}\left[\text{Pr}\bigg[R_{\rm L_{1}}\le \theta^{\frac{1}{\alpha_{\rm L}}}R_{\rm L_{0}},R_{\rm N_{0}} \ge (\frac{\eta_{\rm N}}{\eta_{\rm L}})^{\frac{1}{\alpha_{\rm N}}}R_{\rm L_{1}}^{\frac{\alpha_{\rm L}}{\alpha_{\rm N}}}\bigg]\right].
%&\stackrel{\text{(a)}}{=}\int_{\Delta h_{\rm u}}^{l_{\rm L\_N}}\int_{r_{\rm L_{0}}}^{\theta^{\frac{1}{\alpha_{\rm L}}}r_{\rm L_{0}}}f_{R_{{\rm L}_{0}},R_{{\rm L}_{1}}}(r_{{\rm L}_{0}},r_{{\rm L}_{1}})dr_{\rm L_{1}}dr_{\rm L_{0}}\\
%&+\int_{l_{\rm L\_N}}^{+\infty}\int_{r_{\rm L_{0}}}^{\theta^{\frac{1}{\alpha_{\rm L}}}r_{\rm L_{0}}}f_{R_{{\rm L}_{0}},R_{{\rm L}_{1}}}(r_{{\rm L}_{0}},r_{{\rm L}_{1}})\exp\bigg(-2\pi\lambda_{b}\int_{0}^{l(\theta^{\frac{1}{\alpha_{\rm N}}}d_{\rm L\_N}(r_{\rm L_{0}}))}zp^{\rm N}(z)dz\bigg)dr_{\rm L_{1}}dr_{\rm L_{0}},
	\end{aligned}
\end{equation}
Following the similar steps as of the proof for $\mathcal{A}_{\rm L_{0},L_{1}}$, and considering the range of $r_{\rm L_{0}}$ shown in (\ref{d_rL0}), we complete the proof of $\mathcal{A}_{\rm L_{0}, L_{1}}$ in (\ref{fun21}). Similarly, we can prove $\mathcal{A}_{\rm N_{0},N_{1}} $,  $\mathcal{A}_{\rm L_{0},N_{0}} $, and $\mathcal{A}_{\rm N_{0},L_{0}}$, which are given by (\ref{fun22}), (\ref{fun23}) and (\ref{fun24}), respectively.

\emph{C. Proof of Lemma 4}\\
\indent
%L0的pdf
%We define $f_{\widetilde{R}_{\rm L_{0}}}(r_{\rm L_{0}}) $ as the PDF of the distance between the typical Non-CoMP AU and its serving BS, i.e., $\mathcal{B}=\{b_{\rm  L_{0}}\}$. 
We first derive the CDF of the distance $\widetilde{R}_{\rm L_{0}}$, which is given by
\begin{equation}	
	\label{pdf_rL0}
	\begin{aligned}
		&\text{Pr}[R_{\rm L_{0}} < r_{\rm L_{0}} | \mathcal{B} = \{ b_{\rm  L_{0}}\}]
		= \frac{1}{\mathcal{A}_{\rm  L_{0}}}\text{Pr}\left[R_{\rm L_{0}} < r_{\rm L_{0}},\eta_{\rm L}R_{\rm L_{0}}^{-\alpha_{\rm L}} \ge \theta \eta_{\rm N}R_{\rm N_{0}}^{-\alpha_{\rm N}},R_{\rm L_{0}}^{-\alpha_{\rm L}} \ge \theta  R_{\rm L_{1}}^{-\alpha_{\rm L}}\right] \\
        &\stackrel{\text { (a)}}{=}\frac{1}{\mathcal{A}_{\rm  L_{0}}}\int_{\Delta h_{\rm u}}^{r_{\rm L_{0}}}\text{Pr}\left [R_{\rm L_{1}} \ge \theta^{\frac{1}{\alpha_{\rm L}}}R_{\rm L_{0}}\right ]\text{Pr}\left [R_{\rm N_{0}} \ge \left(\theta\frac{\eta_{\rm N}}{\eta_{\rm L}}\right)^{\frac{1}{\alpha_{\rm N}}}R_{\rm L_{0}}^{\frac{\alpha_{\rm L}}{\alpha_{\rm N}}} \right ] f_{R_{\rm L_{0}}}(r)dr\\
		&\stackrel{\text{(b)}}{=}
\frac{1}{\mathcal{A}_{\rm L_{0}}}\left[\int_{\Delta h_{\rm u}}^{r_{\rm L_{0}}}\int_{\theta^{\frac{1}{\alpha_{\rm L}}}r}^{\infty}\exp\left(-2\pi\lambda_{b} \int_{0}^{l(\widetilde{d}_{\rm L\_N}(r))}zp^{\rm N}(z)dz\right)f_{R_{{\rm L}_{0}},R_{{\rm L}_{1}}}(r,r_{2})dr_{2}dr\right],\\	
	\end{aligned}
\end{equation}
where (a) follows from the CDF of $R_{\rm L_0}$, (b) is due to the null probability of the inhomogeneous PPP $\Phi^{\rm N}_{\rm B}$, and %$l(x)\triangleq \sqrt{x^2-(\Delta h_{\rm u})^2}$,
$\widetilde{d}_{\rm L\_N}(r_{\rm L_{0}})$ is given by (\ref{d_rL0}). %with $d_{\rm L\_N}(r_{\rm L_{0}})\triangleq \left(\dfrac{\eta_{\rm N}}{\eta_{\rm L}}\right)^{\frac{1}{\alpha_{\rm N}}}r_{\rm L_{0}}^{\frac{\alpha_{\rm L}}{\alpha_{\rm N}}}$. 
At last, substituting $f_{R_{\rm L_0},R_{\rm L_1}}(r_{\rm L_0},r_{\rm L_1})$ in (\ref{fun16}) and taking the derivative of the CDF with regards to $r_{\rm L_{0}}$, i.e., $f_{\widetilde{R}_{\rm L_{0}}}(r_{\rm L_{0}})= \frac{\partial F_{\widetilde{R}_{\rm L_{0}}}(r_{\rm L_{0}})} {\partial r_{\rm L_{0}}}$, we complete the proof of
$f_{\widetilde{R}_{\rm L_{0}}}(r_{\rm L_{0}})$ in
(\ref{fun25}). Following the same steps as that of the proof for $f_{\widetilde{R}_{\rm L_{0}}}(r_{\rm L_{0}})$, we can complete the proof of $f_{\widetilde{R}_{\rm N_{0}}}(r_{\rm N_{0}})$, which is given by (\ref{fun26}).

\emph{D. Proof of Lemma 5}\\
\indent
%We first consider the case that the typical CoMP AU is served by the first two closest LoS BSs, i.e., $\mathcal{B}=\{b_{\rm L_{0}},b_{\rm L_{1}}\}$, and define $f_{\widetilde{R}_{\rm L_{0}},\widetilde{R}_{\rm L_{1}}}(r_{\rm L_{0}},r_{\rm L_{1}})$ as the joint PDF of the distances between the typical CoMP AU and the two serving LoS BSs. 
To obtain the joint PDF $f_{\widetilde{R}_{\rm L_{0}},\widetilde{R}_{\rm L_{1}}}(r_{\rm L_{0}},r_{\rm L_{1}})$, we first derive the joint CDF $F_{\widetilde{R}_{\rm L_{0}},R_{\rm L_{1}}}(r_{\rm L_{0}},r_{\rm L_{1}})$, which is given by
\begin{equation}
	\begin{aligned}
 			&\text{Pr}[R_{\rm L_{0}} \le r_{\rm L_{0}},R_{\rm L_{1}} \le r_{\rm L_{1}} | \mathcal{B} = \{ b_{\rm L_{0}}, b_{\rm L_{1}}\}]\\
			&= \frac{1}{\mathcal{A}_{\rm L_{0},L_{1}}}\text{Pr}[R_{\rm  L_{0}} \le r_{\rm L_{0}},R_{\rm  L_{1}} \le r_{\rm L_{1}} ,\theta R_{{\rm  L_{1}}}^{-\alpha_{\rm  L}}\ge R_{{\rm  L_{0}}}^{-\alpha_{\rm  L}},
			\eta_{\rm  L}R_{{\rm  L_{1}}}^{-\alpha_{\rm L}}\ge \eta_{\rm N}R_{{\rm N_{0}}}^{-\alpha_{\rm N}}]\\
			&=\frac{1}{\mathcal{A}_{\rm L_{0},L_{1}}}\int_{\Delta h_{\rm u}}^{r_{\rm L_{0}}}\text{Pr}\left [R_{\rm L_{1}} \ge \theta^{\frac{1}{\alpha_{\rm L}}}R_{\rm L_{0}}\right ]
			\text{Pr}\left [R_{\rm N_{0}} \ge (\theta\frac{\eta_{\rm N}}{\eta_{\rm L}})^{\frac{1}{\alpha_{\rm N}}}R_{\rm L_{1}}^{\frac{\alpha_{\rm L}}{\alpha_{\rm N}}} \right ] f_{R_{\rm L_{0}}}(r)dr\\   
%	\end{flalign}   
% \nonumber
%\end{equation}
%\begin{equation}
%	\label{fun46}
%	\begin{aligned}   
			&\stackrel{\text { (a) }}{=}
            \begin{cases}
				\frac{1}{\mathcal{A}_{\rm  L_{0},L_{1}}}[\int_{\Delta h_{\rm u}}^{r_{\rm L_{0}}}\int_{r}^{r_{\rm L_{1}}}f_{R_{{\rm L}_{0}},R_{{\rm L}_{1}}}(r,r_{2})dr_{2}dr],\qquad \qquad \qquad \qquad \qquad \qquad \quad {\text{if}} 
                \ r_{0} \in [\Delta h_{\rm u}, l_{\rm L\_N}),  \\
				\frac{1}{\mathcal{A}_{\rm L_{0},L_{1}}}[\int_{l_{\rm L\_N}}^{r_{\rm L_{0}}}\int_{r}^{r_{\rm L_{1}}}	
                \exp(-2\pi\lambda_{b}\int_{0}^{l\left(\theta^{\frac{1}{\alpha_{\rm N}}}d_{\rm L\_N}(r_{\rm L_{0}})\right)}zp^{\rm N}(z)dz)f_{R_{{\rm L}_{0}},R_{{\rm L}_{1}}}(r,r_{2})dr_{2}dr], \ {\text{if}} \ r_{0} \ge l_{\rm L\_N},
			\end{cases}
	\end{aligned}
\end{equation}
where (a) follows from the null probability of inhomogenous PPP $\Phi^{\rm N}_{\rm B}$, 
%$l(x)\triangleq \sqrt{x^2-(\Delta h_{\rm u})^2}$, and $d_{\rm L\_N}(r_{\rm L_{0}})\triangleq \left(\dfrac{\eta_{\rm N}}{\eta_{\rm L}}\right)^{\frac{1}{\alpha_{\rm N}}}r_{\rm L_{0}}^{\frac{\alpha_{\rm L}}{\alpha_{\rm N}}}$. 
At last, substituting $f_{R_{\rm L_0},R_{\rm L_1}}(r_{\rm L_0},r_{\rm L_1})$ in (\ref{fun16}) and taking the derivative of the CDF with regards to $r_{\rm L_{0}}$ and $r_{\rm L_{1}}$, i.e., $f_{\widetilde{R}_{\rm L_{0}},\widetilde{R}_{\rm L_{1}}}(r_{\rm L_{0}},r_{\rm L_{1}})= \frac{\partial^{2}F_{\widetilde{R}_{\rm L_{0}},\widetilde{R}_{\rm L_{1}}}(r_{\rm L_{0}},r_{\rm L_{1}}) }{\partial r_{\rm L_{0}} \partial r_{\rm L_{1}}}$, we complete the proof of $f_{\widetilde{R}_{\rm L_{0}},\widetilde{R}_{\rm L_{1}}}(r_{\rm L_{0}},r_{\rm L_{1}})$ in (\ref{fun27}). Following the same steps as that of the proof for $f_{\widetilde{R}_{\rm L_{0}},\widetilde{R}_{\rm L_{1}}}(r_{\rm L_{0}},r_{\rm L_{1}})$, we can complete the proof of
$f_{\widetilde{R}_{\rm N_{0}},\widetilde{R}_{\rm N_{1}}}(r_{\rm N_{0}},r_{\rm N_{1}})$, $f_{\widetilde{R}_{\rm N_{0}},\widetilde{R}_{\rm L_{0}}}(r_{\rm N_{0}},r_{\rm L_{0}})$, and $f_{\widetilde{R}_{\rm L_{0}},\widetilde{R}_{\rm N_{0}}}(r_{\rm L_{0}},r_{\rm N_{0}})$, which are given by (\ref{fun28}), (\ref{fun29}) and (\ref{fun30}), respectively. This completes the proof of Lemma 5.

\emph{E. Proof of Theorem 2}\\
\indent
Referring to (\ref{fun11}), the numerator $\left|\sum_{k=0}^{1}\left(\rho_{\rm u}P_{\rm t}\zeta_{v}(b_{k})\right)^{\frac{1}{2}}\widetilde{\omega}_{k,0}\right|^{2}$ represents the square of a weighted sum of two Nakagami-$m$ RVs. Since closed-form expression is unknown, we use the Cauchy-Schwarz's inequality to obtain the upper bound of $\left|\sum_{k=0}^{1}(\rho_{\rm u}P_{\rm t}\zeta_{v}(b_{k}))^{\frac{1}{2}}\widetilde{\omega}_{k,0}\right|^{2} $ as follows
\begin{equation}
	\label{fun35}
\begin{aligned}
&\left|\sum_{k=0}^{1}\left(\rho_{\rm u}P_{\rm t}\zeta_{v}(b_{k})\right)^{\frac{1}{2}}\widetilde{\omega}_{k,0}\right|^{2} = \left|\sum_{k=0}^{1}\left(\rho_{\rm u}P_{\rm t}\zeta_{v}(b_{k})\right)^{\frac{1}{2}}\frac{\omega_{k0}^{*}}{|\omega_{k0}|}\omega_{k0}\right|^{2} =\rho_{\rm u}P_{\rm t}\left(\sum_{k=0}^{1}Q_{k}\right)^{2} \le 2\rho_{\rm u}P_{\rm t}\left(\sum_{k=0}^{1}Q_{k}^{2}\right),
\end{aligned}
\end{equation}
where $ Q_{k}=(\zeta_{v}(b_{k}))^{\frac{1}{2}}\widetilde{\omega}_{k,0} =(\zeta_{v}(b_{k}))^{\frac{1}{2}}|\omega_{k,0}|$ is a scaled Nakagami-$m$ RV. Since $ \omega_{k,0} \sim $ Nakagami-$m$, according to the scaling property of the Gamma distribution, we have $Q_{k}^{2} \sim \Gamma(\kappa_{k}=m_{v}, \theta_{k} = \zeta_{v}(b_{k})/ m_{v})$. To achieve a tractable statistical equivalent of two Gamma RVs with different scale parameters $\theta_{k}$, we adopt the method of second-order moment matching for Gamma RVs. It is shown that the equivalent Gamma distribution, denoted by $\mathcal{J} \sim \Gamma(K,\Theta)$, has the same first-order and second-order moments with the following parameters
\begin{equation}
	\label{K-Theta}
	\begin{aligned} K=\frac{(\sum_{k=0}^{1}\kappa_{k}\theta_{k})^{2}}{\sum_{k=0}^{1}\kappa_{k}\theta_{k}^{2}}
=\frac{m_{v}(\sum_{k=0}^{1}\zeta_{v}(b_{k}))^{2}}{\sum_{k=0}^{1}(\zeta_{v}(b_{k}))^{2}},
	\Theta = \frac{\sum_{k=0}^{1}\kappa_{k}\theta_{k}^{2}}{\sum_{k=0}^{1}\kappa_{k}\theta_{k}}
=\frac{\sum_{k=0}^{1}m_{v}(\zeta_{v}(b_{k})/ m_{v})^{2}}{\sum_{k=0}^{1}m_{v}\zeta_{v}(b_{k})}.
	\end{aligned}
\end{equation}
\indent
To derive the upper bound of the shape parameter $K$, we consider the following two cases: \\
i) the two cooperative BSs are of the same type, i.e., $ \mathcal{B}=\{b_{\rm L_{0}},b_{\rm L_{1}}\} $ or $ \mathcal{B}=\{b_{\rm N_{0}},b_{\rm N_{1}}\} $; \\
ii) the two cooperative BSs are of different types, i.e., i.e., $ \mathcal{B}=\{b_{\rm L_{0}},b_{\rm N_{0}}\} $ or $ \mathcal{B}=\{b_{\rm N_{0}},b_{\rm L_{0}}\}$. \\
For case i), we use the Cauchy-Schwarz's inequality, which leads to
	$ K = \frac{m_{v}(\sum_{k=0}^{1}\zeta_{v}(b_{k}))^{2}}{\sum_{k=0}^{1}(\zeta_{v}(b_{k}))^{2}} \le 2m_{v}$, $v=\rm L$ or $\rm N$. For case ii), we derive the upper bound of $K$ using the weighted norm inequality as $K \le m_{\rm N} + m_{\rm L}$.\\
\indent
Given $r_{\rm L_{0}}$ and $r_{\rm L_{1}}$, the conditional coverage probability can be derived by
%we first derive the coverage probability for the case $ \mathcal{B}=\{b_{\rm L_0}, b_{\rm L_{1}}\} $, where we define $r_{\rm L_{0}}$ and $r_{\rm L_{1}}$ as the distances between the typical CoMP AU and the serving BS $b_{\rm L_{0}}$ and $b_{\rm L_{1}}$, respectively. 
\begin{equation}
	\label{con-L0L1}
	\begin{aligned}
    &\mathbb P_{{\rm L_{0},L_{1}}| r_{\rm L_{0}},r_{\rm L_{1}}}(T)
    %&= \text{Pr}\left[\varUpsilon_{\rm u}^{\rm C}>{T} \mid r_{\rm L_{0}},r_{\rm  L_{1}}\right]
    %=\text{Pr}\left[\frac{\rho_{\rm u}{P_{\rm t}}\left(\sum_{k=0}^{1}Q_{k}\right)^{2}}{\rho_{\rm t}{P_{\rm t}}\left(\sum_{k=0}^{1}Q_{k}^{2}\right)+I} \ge {T} | r_{\rm L_{0}},r_{\rm L_{1}}\right]\\
	\overset{\text{(a)}}{\le} \text{Pr}\left[\frac{2\rho_{\rm u}{P_{\rm t}}\mathcal{J}}{\rho_{\rm t}P_{\rm t}\mathcal{J}+ I} \ge {T} | r_{\rm L_{0}},r_{\rm L_{1}}\right]
		\overset{\text{(b)}}{\approx} \text{Pr}\left[\mathcal{J}>\frac{T I}{(2\rho_{\rm u}-\rho_{\rm t}{T}){P_{\rm t}}}\right]\\
		&\overset{\text{(c)}}{=}\dfrac{\Gamma\left(K,\dfrac{ T I}{\left(2\rho_{\rm u}-\rho_{\rm t}T\right){P_{\rm t}}\Theta}\right)}{\Gamma(K)}	
        %\overset{\text{(d)}}{=}\mathbb{E}_{I}\left[\sum_{k=0}^{K-1}\left(\frac{TI}{\left(2\rho_{\rm u}-\rho_{t}T\right){P_{\rm t}}\Theta}\right)^{k}\exp\left(-\frac{TI}{\left(2\rho_{\rm u}-\rho_{\rm t}{T}\right){P_{\rm t}}\Theta}\right)\right]\\
        \overset{\text{(d)}}{=}\mathbb{E}_{I}\left[\sum_{k=0}^{K-1}\left(sI)^{k}\exp\left(-sI\right)\right)\right]|_{s=\frac{T}{\left(2 \rho_{\rm u}-\rho_{\rm t}T\right)P_{\rm t}\Theta}}\\
        %&=\sum_{k=0}^{K-1}\frac{\left(\dfrac{-{T}}{(2\rho_{\rm u}-\rho_{\rm t}{T}){P_{\rm t}}\Theta}\right)^{k}}{k!}\mathbb{E}_{I}\left[(-I)^{k}\exp\left(-\frac{TI}{\left(2\rho_{\rm u}-\rho_{\rm t}{T}\right){P_{\rm t}}\Theta}\right)\right]\\		
        &\overset{\text{(e)}}{=}
		\begin{cases}
			\sum_{k=0}^{2 m_{\rm L}-1} \frac{\left(-s\right)^{k}}{k !} \frac{\partial^{k}}{\partial s^{k}} \mathcal{\widehat{L}}_{I}\left(s\right)|_{s=\frac{T}{\left(2 \rho_{\rm u}-\rho_{\rm t}T\right)P_{\rm t}\Theta}}, &{\text{if}}\  r_{\rm L_{0}} \in [\Delta h_{\rm u}, l_{\rm L\_N}),  \\
				\sum_{k=0}^{2 m_{\rm L}-1} \frac{\left(-s\right)^{k}}{k !} \frac{\partial^{k}}{\partial s^{k}} \mathcal{\widetilde{L}}_{I}\left(s\right)|_{s=\frac{T}{\left(2 \rho_{\rm u}-\rho_{\rm t}T\right)P_{\rm t}\Theta}},&{\text{if}\    r_{\rm L_{0}} \ge l_{\rm L\_N}},
		\end{cases}
	\end{aligned}
\end{equation}
where 
%$l_{\rm L\_N}\triangleq\left(\dfrac{\eta_{\rm L}}{\eta_{\rm N}}\right)^{\frac{1}{\alpha_{\rm L}}}(\Delta h_{\rm u})^{\frac{\alpha_{\rm N}}{\alpha_{\rm L}}}$, 
$I\triangleq{\sum_{i\in\Phi_{\rm B}\backslash{\{b_{\rm L_0}, b_{\rm L_1}\}}}P_{\rm t}\zeta_{v}(b_{i})|\widetilde{\omega}_{i,0}|^2}$, $ \Theta =\frac{\sum_{k=0}^{1}(\zeta_{\rm L}(b_{{\rm L}_{k}}))^2}{m_{\rm L}\sum_{k=0}^{1}\zeta_{\rm L}(b_{{\rm L}_k})}$, (a) follows from the Cauchy-Schwarz's inequality, (b) follows from the Gamma approximation by rounding the shape parameter $K = 2 m_{\rm L}$, and (c) is due to the fact that for a Gamma-distributed RV $ Z \sim \Gamma[k_{z}, \theta_{z}]$ with integer $k_{z}$, we have 
%$ P[Z > x] = \dfrac{\Gamma(k_{z},\frac{x}{\theta_{Z}})}{\Gamma(k_{z})} $. 
%As a result, for an independent RV $X$, we have $\text{Pr}[Z > X] =\mathbb{E}_{X}\left[\dfrac{\Gamma(k_{z},\frac{X}{\theta_{z}})}{\Gamma(k_{z})}\right]$.
%By using the fact that 
$ \text{Pr}[Z > x] =\dfrac{\Gamma(k_{z},\frac{y}{\theta_{z}})}{\Gamma(k_{z})} = \sum_{k=0}^{k_{z}-1}\frac{y^{k}\exp(-y)}{k!}$, which leads to (d). Finally, with $\frac{\partial^{k}}{\partial{k_{z}}}[\exp(-z Y)] = (-Y)^{k}\exp(-zY)$, we derive (e), where the Laplace transform $\mathcal{\widehat{L}}_{\rm I}(s)$ and $\mathcal{\widetilde{L}}_{\rm I}(s)$ are given by (\ref{laplace-LL-1}) and (\ref{laplace-LL-2}), respectively. 
\begin{comment}
\begin{equation}
	\begin{aligned}
		\label{Lap-L0L1}
        \mathcal{\widehat{L}}_{\rm I}(s)
        =\exp\biggl(&-2\pi\lambda_{b}\int_{0}^{+\infty}\left[1-(\dfrac{m_{\rm N}}{m_{\rm N}+s\eta_{\rm N}(\sqrt{z^2+\Delta h_{\rm u}^{2}})^{-\alpha_{\rm N}}})^{m_{\rm N}}\right]zp^{\rm N}(z)dz\\
		&-2\pi\lambda_{b}\int_{l(r_{\rm L_{1}})}^{+\infty}\left[1-\left(\dfrac{m_{\rm L}}{m_{\rm L}+s\eta_{\rm L}(\sqrt{z^2+\Delta h_{\rm u}^{2}})^{-\alpha_{\rm L}}}\right)^{m_{\rm L}}\right]zp^{\rm L}(z)dz\biggl),\\
	\end{aligned}	
\end{equation}
\begin{equation}
	\begin{aligned}
		\label{laplace}
        \mathcal{\widetilde{L}}_{\rm I}(s)
        =\exp\bigg(&-2\pi\lambda_{b}\int_{l(d_{\rm L\_N}(r_{\rm L_{1}}))}^{+\infty}
        \left[1-(\dfrac{m_{\rm N}}{m_{\rm N}+s\eta_{\rm N}(\sqrt{z^2+\Delta h_{\rm u}^{2}})^{-\alpha_{\rm N}}})^{m_{\rm N}}\right]zp^{\rm N}(z)dz\\
		&-2\pi\lambda_{b}\int_{l(r_{\rm L_{1}})}^{+\infty}\left[1-\left(\dfrac{m_{\rm L}}{m_{\rm L}+s\eta_{\rm L}(\sqrt{z^2+\Delta h_{\rm u}^{2}})^{-\alpha_{\rm L}}}\right)^{m_{\rm L}}\right]zp^{\rm L}(z)dz\bigg),\\
	\end{aligned}	
\end{equation}
\end{comment}
%where $l(x)=\sqrt{x^{2}-\Delta h_{\rm u}^{2}}$, $d_{\rm L\_N}(r_{\rm L_{1}})=\left(\dfrac{\eta_{\rm N}}{\eta_{\rm L}}\right)^{\frac{1}{\alpha_{\rm N}}}{r_{\rm L_{1}}^{\frac{\alpha_{\rm L}}{\alpha_{\rm N}}}}$. 
Finally, by averaging over $r_{\rm L_{1}}$ and $r_{\rm L_{0}}$, we complete the proof of the coverage probability when the typical CoMP AU associating with $\mathcal{B}=\{b_{\rm L_{0}}, b_{\rm L_{1}}\}$ in (\ref{Cov-L0L1}) of Theorem 2. Following the same steps as that for the proof of $\mathbb P_{\rm L_{0},L_{1}}(T)$, we complete the proof of $\mathbb P_{\rm N_{0},N_{1}}(T)$, $\mathbb P_{\rm L_{0},N_{0}}(T)$, and $\mathbb P_{\rm N_{0},L_{0}}(T)$ in (\ref{Cov-N0N1}), (\ref{Cov_L0N0}), and (\ref{Cov_N0L0}), respectively. This completes the proof of Theorem 2.\\
\indent

\footnotesize

\bibliographystyle{IEEEtran}
\bibliography{author}
	
\end{document}